\documentclass[
  journal=largetwo,
  manuscript=article-type,
  year=2024,
  volume=37,
]{cup-journal}

\usepackage{amsmath}
\usepackage{amsfonts}
\usepackage{amssymb}
\usepackage[nopatch]{microtype}
\usepackage{booktabs}
\usepackage[table]{xcolor}
\usepackage{aas_macros}

\defcitealias{NFW1995}{NFW 1995}
\defcitealias{NFW1996}{NFW 1996}
\defcitealias{NFW1997}{NFW 1997}

\usepackage{soul}

\title{Predicting the Non-Thermal Pressure in Galaxy Clusters}

\author{A. Sullivan}
\affiliation{International Centre for Radio Astronomy Research, The University of Western Australia, 35 Stirling Highway, Crawley, Western Australia, 6009, Australia}
\alsoaffiliation{ARC Centre of Excellence for All Sky Astrophysics in 3 Dimensions (ASTRO 3D)}

\author{S. Shabala}
\affiliation{School of Natural Sciences, University of Tasmania, Hobart, Tasmania, Australia}

\author{C. Power}
\affiliation{International Centre for Radio Astronomy Research, The University of Western Australia, 35 Stirling Highway, Crawley, Western Australia, 6009, Australia}
\alsoaffiliation{ARC Centre of Excellence for All Sky Astrophysics in 3 Dimensions (ASTRO 3D)}

\author{C. Bottrell}
\affiliation{International Centre for Radio Astronomy Research, The University of Western Australia, 35 Stirling Highway, Crawley, Western Australia, 6009, Australia}

\author{A. Robotham}
\affiliation{International Centre for Radio Astronomy Research, The University of Western Australia, 35 Stirling Highway, Crawley, Western Australia, 6009, Australia}
\alsoaffiliation{ARC Centre of Excellence for All Sky Astrophysics in 3 Dimensions (ASTRO 3D)}

\email[A. Sullivan]{andrew.sullivan@icrar.org}

\keywords{methods: analytical - cosmology: dark matter - galaxies: clusters: general - galaxies: clusters: intracluster medium - X-rays: galaxies: clusters} 

\begin{document}

\begin{abstract}
We investigate the relationship between a galaxy cluster's hydrostatic equilibrium state, the entropy profile, $K$, of the intracluster gas, and the system's non-thermal pressure (NTP), within an analytic model of cluster structures. When NTP is neglected from the cluster's hydrostatic state, we find that the gas' logarithmic entropy slope, $k\equiv \mathrm{d}\ln K/\mathrm{d}\ln r$, converges at large halocentric radius, $r$, to a value that is systematically higher than the value $k\simeq1.1$ that is found in observations and simulations. By applying a constraint on these `pristine equilibrium' slopes, $k_\mathrm{eq}$, we are able to predict the required NTP that must be introduced into the hydrostatic state of the cluster. We solve for the fraction, $\mathcal{F}\equiv p_\mathrm{nt}/p$, of NTP, $p_\mathrm{nt}$, to total pressure, $p$, of the cluster, and we find $\mathcal{F}(r)$ to be an increasing function of halocentric radius, $r$, that can be parameterised by its value in the cluster's core, $\mathcal{F}_0$, with this prediction able to be fit to the functional form proposed in numerical simulations. The minimum NTP fraction, as the solution with zero NTP in the core, $\mathcal{F}_0=0$, we find to be in excellent agreement with the mean NTP predicted in non-radiative simulations, beyond halocentric radii of $r\gtrsim0.7r_{500}$, and in tension with observational constraints derived at similar radii. For this minimum NTP profile, we predict $\mathcal{F}\simeq0.20$ at $r_{500}$, and $\mathcal{F}\simeq0.34$ at $2r_{500}$; this amount of NTP leads to a hydrostatic bias of $b\simeq0.12$ in the cluster mass $M_{500}$ when measured within $r_{500}$. Our results suggest that the NTP of galaxy clusters contributes a significant amount to their hydrostatic state near the virial radius, and must be accounted for when estimating the cluster's halo mass using hydrostatic equilibrium approaches.
\end{abstract}

\section{1. Introduction}

Galaxy clusters are the largest gravitationally bound structures in the universe, and are important astrophysical environments for understanding the interplay between dark matter halos and their hot gaseous atmospheres. The hot, ionised component of the intracluster gas is observed via its X-ray emission, which is expected to scale with the galaxy cluster's underlying dark matter halo mass. 

Modern high precision X-ray telescopes such as XMM-\textit{Newton} \autocite[e.g.][]{XMMNewton}, \textit{Chandra} \autocite[e.g.][]{Chandra} and eROSITA \autocite[e.g.][]{eROSITA} have enabled precise fits to be made for the radial profile of the intracluster gas density, temperature, and pressure, which can be related to the cluster's halo mass through the assumption of hydrostatic equilibrium. These hydrostatic halo masses can be correlated with observable probes of the intracluster gas emission to produce a scaling relation --- typically to either a mean-weighted X-ray temperature \autocite[e.g.][]{Vikhlinin2006, Vikhlinin2009, Babyk2023}, or the shift in the Cosmic Microwave Background (CMB) known as the Sunyaev-Zeldovich \autocite[SZ;][]{SZ1970, SZ1972} effect, arising from photon interactions with energetic electrons in the intracluster gas \autocite[e.g.][]{Vanderlinde2010, Andersson2011}.

In this approach, the halo mass is recovered up to a hydrostatic bias, which is believed to lead to an underestimate in the halo mass in relaxed galaxy clusters by at least $\sim 10-20\%$ \autocite{Martizzi2016, Ettori2022}. This hydrostatic bias is attributed to the non-thermal pressure (NTP, hereafter) contributing to the cluster's hydrostatic state, which is neglected when calculating hydrostatic halo masses. NTP is defined as the pressure of a system that is not attributed to the random motion of the intracluster gas; NTP will be produced by shocks, mergers and feedback processes. 

One of the biggest challenges in estimating halo masses this way is accurately quantifying the hydrostatic bias, which relies on quantifying the fraction of NTP to total pressure in a cluster, at any given halocentric radius. Observationally, the NTP fraction in galaxy clusters is constrained to be $\lesssim 11\% $ at halocentric radii of $r_{500}$ for systems with similar mass and at similar redshifts; this is obtained by comparing hydrostatic halo masses with halo masses computed from gravitational lensing \autocite{Siegel2018}. Other observational studies have predicted NTP fractions of $\lesssim 9\%$ and $\lesssim 15\%$ at halo radii of $r_{500}$ and $r_{200}$, respectively, when calibrating hydrostatic gas mass fractions to the expected universal gas fraction \autocite{Eckert2019}; \textcolor{black}{whilst the NTP fraction inferred from X-ray surface brightness fluctuations has been predicted at $\sim 7\%$ near $r_{500}$, and only $\sim 1-2\%$ closer toward the cluster's core \autocite{Dupourque2023}. This latter constraint is in relatively good agreement to precise modelling by} the \textit{Hitomi} satellite, which by directly measuring the turbulent gas motion within the central 100 $\mathrm{kpc}$ of the Perseus galaxy cluster, has constrained the fraction of kinetic to thermal pressure support to be $\sim 2-7\%$, with an upper bound of at most $\sim 11-13\%$ \autocite{Hitomi2016, Hitomi2018}. 
These observational constraints suggest that galaxy clusters are consistent with little or no NTP in their central core, and a radially increasing NTP fraction that is not expected to be more than $\sim 10\%$ at $r_{500}$. 

The importance of NTP for hydrostatic halo mass estimation has motivated its study in state-of-the-art hydrodynamic cosmological simulations of galaxy clusters. In general, non-radiative simulations predict the cluster's NTP at a factor of $\sim 3$ above observational constraints, with a NTP fraction of $\sim 20 - 40\%$ near $r_{200}$ \autocite[e.g.][]{Nelson2014, Martizzi2016}. These predictions are consistent across studies that vary the sub-grid physics in radiative hydrodynamic codes \autocite[e.g.][]{Pearce2020}. Related work has shown that these numerical constraints on the NTP will vary when defined in terms of different definitions of gas motion --- total random motion, turbulent motion, radial motion, or any combination of these --- and that each definition will vary in its contribution to the NTP fraction associated with the hydrostatic bias \autocite[see][]{Angelinelli2020}. 

Interestingly, recent observational constraints on the NTP of galaxy clusters with gravitational lensing observations have produced results that are consistent with numerical predictions at $95\%$ confidence, predicting a radially increasing profile with large variation, consistent with a NTP fraction of $\sim 20\%$ near $r_{200}$ \autocite{Sayers2021}, which is in tension with other observational constraints. Unfortunately, this lack of consensus between simulations and observations, and between different sets of observations, represents an important limit in the utility of hydrostatic masses as a tool for halo mass estimation. 

In contrast to NTP, the gas entropy, $K$, is a thermodynamic property of galaxy clusters that is well constrained across both simulations and observations. Astrophysical entropy is related to, but distinct, from statistical entropy from thermodynamics. In galaxy clusters, the gas entropy is a tracer of the evolution of the intracluster gas phase, as it is a sensitive probe of non-gravitational processes, and hence is a strong indicator of the thermal state of the cluster. In particular, $K$ is known to scale with the cluster's halocentric radius, $r$, as a broken radial power law, scaling differently inside and outside the influence of radiative heating and non-gravitational feedback. In simulations, where the hot gaseous atmosphere is shaped by non-radiative, gravitational processes, the gas entropy is found to follow the radial scaling $K(r) \propto r^{1.1}$ out to $r \simeq r_{200}$ \autocite{TozziNorman2001, Voit2005}. This is consistent with observational fits, recovering this power law slope of $1.1$ beyond cluster radii of $r \gtrsim 0.6 r_{500}$ \autocite{Hogan2017, Ghirardini2019}. Within the region $0.04 r_{500} \lesssim r \lesssim 0.4r_{500}$, observational fits find a gradual increase in the entropy slope with increasing cluster radius, with the entropy slope better fit by a shallower power law, of $K(r) \propto r^{1.05}$, inside this range \autocite{Babyk2018}. Departures from this power law are expected below a radial break of $r \simeq 0.03 r_{500}$ \autocite{Babyk2018}, where non-gravitational processes become increasingly important toward the cluster's central region.

In the central regions of galaxy clusters, the thermal properties of the intracluster gas are often used to classify clusters as either `cool core' (CC) or `non-cool core' (NCC) clusters. Generally, CCs are associated with a temperature drop in the central region, whilst NCCs show a constant or increasing temperature toward the cluster's centre \autocite[for an overview in defining these classifications, see][]{Hudson2010}. In terms of their central gas entropy, CCs are expected to follow a shallower radial power law, scaling as $K(r) \propto r^{2/3}$, as constrained observationally \autocite[e.g.][]{Pagagoulia2014, Hogan2017, Babyk2018, Ghirardini2019}, whereas NCCs are expected to be better described by some `entropy floor', $K(r) \simeq K_0$, in the core. This general understanding and consensus for the scaling of the gas entropy, within and beyond the central region, for both CC or NCC clusters, motivates our use of these expected constraints in modelling the thermal state of galaxy clusters below; in doing so, we can reveal the amount of NTP that is required to maintain hydrostatic equilibrium.  

In previous work \citep[cf.][hereafter S24b]{Sullivan2024b}, we developed an analytic model for galaxy clusters and the properties of their intracluster gas emission, for clusters in virial and hydrostatic equilibrium, and parameterised by the structure and composition of the hot gas and dark matter constituents. However, one important caveat in that model was the lack of NTP, implying an overestimate in its temperature and pressure profiles due to the hydrostatic bias. In this work, we apply the observational and simulation constraints to the gas entropy predicted in that model, and analytically predict the functional form of the required NTP to attain the expected scaling. This allows us to propose an analytic profile for the NTP fraction of galaxy clusters, facilitating a comparison to be made to both the observational and numerical predictions for its radial profile. 

Our general approach, detailing the mathematical connection between the NTP fraction and the gas entropy, is detailed in Section 2. In Section 3, we analyse the gas entropy predicted by our previous model, and propose a weighting function that constrains these profiles to attain the entropy scaling that is expected from the literature. In Section 4 we present the required NTP fraction for our cluster model, and we show how incorporating this profile improves our predictions for the gas' entropy, temperature and thermal pressure profiles. We also comment on the expected hydrostatic bias, and the impact this has on the cluster scaling relations. We present our conclusions in Section 5.

\section{2. Theoretical background and methods}
\vspace{2mm}
\subsubsection{Hydrostatic equilibrium with non-thermal pressure}

In galaxy clusters, the total pressure of the system will comprise the thermal pressure, $p_\mathrm{th}$, exerted by the intracluster gas, as well as the NTP, $p_\mathrm{nt}$, arising due to gravitational shocks and mergers, or due to feedback processes (e.g. powerful outflows driven by active galactic nuclei) in the central regions. For galaxy clusters in hydrostatic equilibrium, the acceleration exerted by this pressure on the intracluster gas will be balanced by the gravitational force generated by the cluster's mass, at any halo radius. 

In the idealised case of a spherically symmetric cluster, the hydrostatic equilibrium condition at any halocentric radius, $r$, is:
\begin{equation}\label{perturbed equilibrium}
    \frac{\mathrm{d} }{\mathrm{d}r}\left[p_\mathrm{th}(r) + p_\mathrm{nt}(r)\right] = -  \rho_\mathrm{gas}(r)\frac{G M(r)}{r^2},
\end{equation}
in terms of the radial derivatives of the thermal pressure profile, $p_\mathrm{th}(r)$, and the NTP profile, $p_\mathrm{nt}(r)$; the halo's enclosed mass, $M(r)$; the density profile of the intracluster gas, $\rho_\mathrm{gas}(r)$; and the gravitational constant, $G$. \textcolor{black}{We note that this assumption of spherical symmetry is not necessary always true for a large population of real clusters \autocite[see, e.g.][]{Campitiello2022}; however, we will assume this holds hereafter. }

In the simplest case, considering only gas and dark matter within the galaxy cluster, the enclosed halo mass, $M(r)$, is given by integrating the sum of the density profiles for the dark matter halo, $\rho_\mathrm{dm}(r)$, and the intracluster gas, $\rho_\mathrm{gas}(r)$, within the spherical volume of radius $r$, as:
\begin{equation}\label{total mass (dark matter + baryonic gas)}
     M(r) = 4\pi \int_0 ^r \left[ \rho_\mathrm{dm}(r^\prime) + \rho_\mathrm{gas}(r^\prime) \right] r^\prime{} ^2 \mathrm{d}r^\prime.
\end{equation}
To solve for the temperature profile, $T(r)$, of the intracluster gas, the thermal pressure profile must be related to the gas' state variables, which for an ideal gas, obeys the relation:
\begin{equation}\label{thermal pressure}
    p_\mathrm{th} = \frac{k_\mathrm{B}T}{\mu m_\mathrm{p}} \rho_\mathrm{gas},
\end{equation}
where $k_\mathrm{B}$ is the Boltzmann constant, $\mu$ is the mean molecular weight, and $m_\mathrm{p}$ is the proton mass. Subsequently, Equation \eqref{perturbed equilibrium} can be expressed as:
\begin{equation}\label{perturbed equilibrium 2}
    \frac{\mathrm{d}}{\mathrm{d}r}\left[ \rho_\mathrm{gas}(r) T(r) + \frac{\mu m_\mathrm{p}}{k_\mathrm{B}} p_\mathrm{nt} (r)\right] = - \frac{G\mu m_\mathrm{p}}{k_\mathrm{B}} \frac{\rho_\mathrm{gas}(r) M(r)}{r^2},
\end{equation}
which can be solved for $T(r)$, given some radial parameterisation for the NTP profile, $p_\mathrm{nt}(r)$. 

When observationally estimating the cluster mass, the NTP term in Equation \eqref{perturbed equilibrium 2} is generally assumed to be zero \citep[e.g.][]{Vikhlinin2006,Vikhlinin2009}, circumventing the need to assume the form of $p_\mathrm{nt}(r)$, which is not well constrained. We took this approach (i.e. neglecting the contribution of $p_\mathrm{nt}$) in S24b. Without a NTP term, the hydrostatic equilibrium state of the cluster is assumed to be entirely balanced by the gas' thermal pressure; this requires the gas to be hotter than it would otherwise be if NTP was present. In this study, hereafter, we will refer to the hydrostatic state without NTP as `pristine equilibrium', to differentiate from the `real equilibrium' state that will include NTP. 

The pristine equilibrium temperature of the gas, which we denote $T_\mathrm{eq}(r)$, is then given by the general solution:
\begin{equation}\label{pristine temperature solution}
    T_\mathrm{eq}(r) = \frac{G\mu m_\mathrm{p}}{k_\mathrm{B}} \frac{1}{\rho_\mathrm{gas}(r)} \int _r^\infty \frac{M(r^\prime) \rho_\mathrm{gas}(r^\prime) \mathrm{d}r^\prime}{r^\prime{} ^2}.
\end{equation}
When including NTP in the cluster's hydrostatic equilibrium state, i.e. Equation \eqref{perturbed equilibrium 2}, the real equilibrium gas temperature, $T(r)$, will instead be given by:
\begin{equation}\label{temperature - pristine equilibrium temperature}
    T(r) = T_\mathrm{eq}(r) \left[1 - \mathcal{F}(r)\right],
\end{equation}
where $\mathcal{F}(r)$ parameterises the fraction of NTP to total pressure in the system:
\begin{equation}\label{NTP definition}
    \mathcal{F}(r) \equiv \frac{p_\mathrm{nt}}{p}(r).
\end{equation}
By this definition, the gas' thermal pressure, $p_\mathrm{th}(r)$, will be related to the NTP fraction, $\mathcal{F}(r)$, by:
\begin{equation}\label{gas thermal pressure}
    p_\mathrm{th}(r) = p(r) \left[1 - \mathcal{F}(r) \right],
\end{equation}
where $p(r)$ is the cluster's total pressure. We assume this total pressure will always be given by the hydrostatic equilibrium condition, Equation \eqref{perturbed equilibrium}, defining the equilibrium pressure:
\begin{equation}\label{pristine pressure solution}
    p_\mathrm{eq}(r) = G \int _r^\infty \frac{M(r^\prime) \rho_\mathrm{gas}(r^\prime) \mathrm{d}r^\prime}{r^\prime{} ^2},
\end{equation}
which, in the pristine equilibrium assumption, will also be the thermal pressure of the gas.

\subsubsection{The gas entropy}

The definition of the intracluster gas entropy, $K$, is:
\begin{equation}\label{entropy definition}
    K \equiv \frac{k_\mathrm{B}T}{n_\mathrm{e}^{2/3}},
\end{equation}
in terms of the Boltzmann constant, $k_\mathrm{B}$, the gas temperature, $T$, and the electron number density, $n_\mathrm{e}$, which is given by:
\begin{equation}\label{electron number density}
    n_\mathrm{e} = \frac{\rho_\mathrm{gas}}{\mu_\mathrm{e} m_\mathrm{p}};
\end{equation}
here $\mu_\mathrm{e}$ is the mean molecular weight of electrons and $m_\mathrm{p}$ is the proton mass. For a spherically symmetric cluster, the radial gas entropy profile, $K(r)$, is then:
\begin{equation}\label{entropy definition in terms of gas variables}
    K(r) = \left[\frac{\mu_\mathrm{e} m_\mathrm{p}}{\rho_\mathrm{gas}(r)}\right]^{2/3}k_\mathrm{B}T(r).
\end{equation}
We can assign a pristine equilibrium gas entropy, $K_\mathrm{eq}(r)$, to a cluster that is in pristine equilibrium, which will be defined as:
\begin{equation}\label{pristine entropy definition}
    K_\mathrm{eq}(r) = \left[\frac{\mu_\mathrm{e} m_\mathrm{p}}{\rho_\mathrm{gas}(r)}\right]^{2/3}k_\mathrm{B}T_\mathrm{eq}(r),
\end{equation}
in terms of the pristine equilibrium temperature of the gas, $T_\mathrm{eq}(r)$.

\subsubsection{The gas entropy slope}

By taking the logarithmic derivative of Equation \eqref{entropy definition in terms of gas variables} with respect to the halocentric radius, $r$, we define the `entropy slope', $k(r)$, in terms of the logarithmic derivatives of the gas' temperature and density, as:
\begin{equation}\label{entropy slope}
    k(r) \equiv \frac{\mathrm{d} \ln K(r)}{\mathrm{d} \ln r} = \frac{\mathrm{d} \ln T(r)}{\mathrm{d} \ln r} - \frac{2}{3} \frac{\mathrm{d} \ln \rho_\mathrm{gas}(r)}{\mathrm{d} \ln r}.
\end{equation}
For a cluster in pristine equilibrium, the associated pristine equilibrium entropy slope, $k_\mathrm{eq}(r)$, is obtained from Equation \eqref{pristine entropy definition}, as:
\begin{equation}\label{pristine equilibrium entropy slope}
    k_\mathrm{eq}(r) \equiv \frac{\mathrm{d} \ln K_\mathrm{eq}(r)}{\mathrm{d} \ln r} = \frac{\mathrm{d} \ln T_\mathrm{eq}(r)}{\mathrm{d} \ln r} - \frac{2}{3} \frac{\mathrm{d} \ln \rho_\mathrm{gas}(r)}{\mathrm{d} \ln r}.
\end{equation}
By relating the gas' real equilibrium temperature, $T(r)$, to its pristine equilibrium temperature, $T_\mathrm{eq}(r)$, by Equation \eqref{temperature - pristine equilibrium temperature}, these entropy slopes can be related to the NTP fraction, $\mathcal{F}(r)$, via the differential equation:
\begin{equation}\label{entropy slope - NTP profile}
    k(r) =  k_\mathrm{eq}(r) + \frac{\mathrm{d} \ln \left[1 - \mathcal{F}(r)\right]}{\mathrm{d} \ln r}.
\end{equation}
The NTP fraction is thus constrained if both $k(r)$ and $k_\mathrm{eq}(r)$ are known.


\begin{table*}[h!]
\begin{tabular}{ |p{2cm}||p{2.2cm}|p{3cm}|p{2.8cm}|p{2.2cm}|p{3cm}| }
\hline
 &  \multicolumn{1}{|c|}{$\boldsymbol{c}$} & \multicolumn{1}{|c|}{$\boldsymbol{\alpha}$} & \multicolumn{1}{|c|}{$\boldsymbol{\eta}$} & \multicolumn{1}{|c|}{$\boldsymbol{d}$} & \multicolumn{1}{|c|}{$\boldsymbol{\varepsilon}$} \\ [1ex]
\hline\hline
\rowcolor{white} \cellcolor{lightgray!30} Definition: & \textit{Concentration} & \textit{Inner density slope of the dark matter profile} & \textit{Fraction of cosmological baryon content}  & \textit{Dilution} & \textit{Inner density slope of the intracluster gas profile} \\ [3ex]
\hline
\rowcolor{white} \cellcolor{lightgray!30} Physical values: &  $ c=2.5$ & $\alpha \in [0, 1.5]$ & $\eta \in [0.6, 1]$ & $ d=1$ & $\varepsilon \in [0, 1]$ \\ [3ex]
\hline
\end{tabular}
\caption{Summary of the five parameters in the ideal baryonic cluster halo model: their symbol, definition and physical values when $\Delta=500$.}
\label{Parameter table}
\end{table*}

\subsubsection{Constraining the non-thermal pressure fraction}

One approach for solving Equation \eqref{entropy slope - NTP profile} is to relate the real entropy slope to its pristine equilibrium value via a weighting function, $w(r)$, such that:
\begin{equation}\label{weighting the entropy slope}
    k(r) = w(r) \cdot k_\mathrm{eq}(r).
\end{equation}
We choose the weighting function such that $k(r)$ matches literature values for the entropy slope over an appropriate range of halocentric radii. Given some form for this weighting function, $w(r)$, we solve Equation \eqref{entropy slope - NTP profile} in the form:
\begin{equation}\label{entropy slope - NTP profile, with weighting}
    k_\mathrm{eq}(r)\left[w(r) - 1\right] =  \frac{\mathrm{d} \mathcal{F}(r)}{\mathrm{d}r}\frac{r}{\left[\mathcal{F}(r) - 1\right]},
\end{equation}
which requires a boundary condition on $\mathcal{F}(r)$. We introduce the parameter $\mathcal{F}_0 \equiv \mathcal{F}(r=0)$, as the cluster's central NTP fraction, such that Equation \eqref{entropy slope - NTP profile, with weighting} can be integrated to give:
\begin{equation}\label{general solution for the NTP}
    \mathcal{F}(r) = 1 + (\mathcal{F}_{0} - 1) \cdot \mathrm{e} ^{\int_{0}^r k_\mathrm{eq}(r^\prime) \left[w(r^\prime) - 1\right]\frac{\mathrm{d}r^\prime}{r^\prime}}.
\end{equation}

\subsubsection{A scale-free approach}


We define the cluster's virial mass, $M_\mathrm{vir}$, in terms of its virial radius, $r_\mathrm{vir}$, such that:

\begin{equation}\label{virial mass definition}
    M_\mathrm{vir} \equiv \frac{4}{3} \pi r_\mathrm{vir}^3 \Delta \rho_\mathrm{crit,0};
\end{equation}
$M_\mathrm{vir}$ is the mass enclosing an average density of $\Delta$ times the present-day critical density of the universe, $\rho_\mathrm{crit,0}$, with the convention $\Delta=500$ usually assumed in studies of galaxy clusters. We therefore use $M_{500}$ as the virial mass and $r_{500}$ as the virial radius. We then define a scale-free dimensionless halocentric radius, $s$, as:
\begin{equation}\label{dimensionless halo radius}
    s \equiv \frac{r}{r_\mathrm{vir}},
\end{equation}
where $r_\mathrm{vir}$ depends on this choice of $\Delta$. 

In terms of $s$, the NTP fraction solution in Equation \eqref{general solution for the NTP} can be expressed as:
\begin{equation}\label{scale-free general solution for the NTP}
    \mathcal{F}(s) = 1 + (\mathcal{F}_{0} - 1) \cdot \mathrm{e} ^{\int_{0}^s k_\mathrm{eq}(s^\prime) \left[w(s^\prime) - 1\right]\frac{\mathrm{d}s^\prime}{s^\prime}}.
\end{equation}
This can be solved, given a scale-free profile for the pristine equilibrium gas entropy slope, $k_\mathrm{eq}(s)$; a scale-free weighting function, $w(s)$; and a prescription for the cluster's central NTP fraction, $\mathcal{F}_0$.

\subsubsection{The ideal baryonic cluster halo profiles}

In general, a model for a cluster's pristine equilibrium entropy slope, $k_\mathrm{eq}(s)$, in scale-free form requires a scale-free structural parameterisation for a cluster's intracluster gas and dark matter halo, to solve for its hydrostatic state. We use the analytic model derived in S24b, which we briefly summarise. 

We obtained an `ideal baryonic cluster halo' in S24b in terms of scale-free density profiles for the dark matter halo, $\rho_\mathrm{dm}(s)$, and the intracluster gas, $\rho_\mathrm{gas}(s)$. For the dark matter, this profile was taken as a generalisation to the NFW \autocite{NFW1995, NFW1996, NFW1997} profile:
\begin{equation}\label{ideal physical baryonic halo profile - dark matter}
    \frac{\rho_\mathrm{dm} (s, c, \alpha, \eta)}{\Delta \rho_\mathrm{crit, 0}} = \frac{(1  - \eta f_\mathrm{b, cos}) u(c, \alpha)}{3s^\alpha (1 + cs)^{3-\alpha}},
\end{equation}
and for the intracluster gas, by the similarly generalised profile: 
\begin{equation}\label{ideal physical baryonic halo profile - baryonic gas}
    \frac{\rho_\mathrm{gas}(s, c, \alpha, \eta, d, \varepsilon)}{\Delta \rho_\mathrm{crit,0}} = \frac{\eta f_\mathrm{b, cos}\mathcal{U}(c, \alpha, d, \varepsilon)}{3s^\varepsilon [1 + \mathcal{C}(c, \alpha, d, \varepsilon) s ]^{3-\varepsilon}}.
\end{equation}
These density profiles are each a function of the dimensionless halocentric radius, $s$, and taken in a dimensionless ratio to some overdensity, $\Delta$, times the present-day critical density of the universe, $\rho_\mathrm{crit,0}$. The five parameters that specify these density profiles are summarised in Table \ref{Parameter table}, along with their recommended value or range in values (cf. S24b). The parameter functions in Equations \eqref{ideal physical baryonic halo profile - dark matter} and \eqref{ideal physical baryonic halo profile - baryonic gas} are then specified in terms of these parameters:
\begin{equation}\label{ideal physical halo concentration function}
    u(c, \alpha) \equiv \left[\int _0 ^1 \frac{s^{2-\alpha} \mathrm{d}s}{(1 + cs)^{3-\alpha}}\right]^{-1},
\end{equation}
\begin{equation}\label{baryon concentration parameter}
    \mathcal{C}(c, \alpha, d, \varepsilon) \equiv \frac{d(\alpha - \varepsilon) + c(3 - \varepsilon)}{3-\alpha},
\end{equation}
and:
\begin{equation}\label{baryon concentration function}
	\mathcal{U}(c, \alpha, d, \varepsilon) \equiv \left[\int _0 ^{1} \frac{s^{2-\varepsilon}\mathrm{d}s}{[1 + \mathcal{C}(c, \alpha, d, \varepsilon) s]^{3-\varepsilon}} \right] ^{-1}.
\end{equation}
We adopt a cosmological baryon fraction of $f_\mathrm{b, cos}=0.158$ \autocite{Planck2016}. 



\begin{figure*}[h!]
    \centering
    \includegraphics[width=\textwidth]{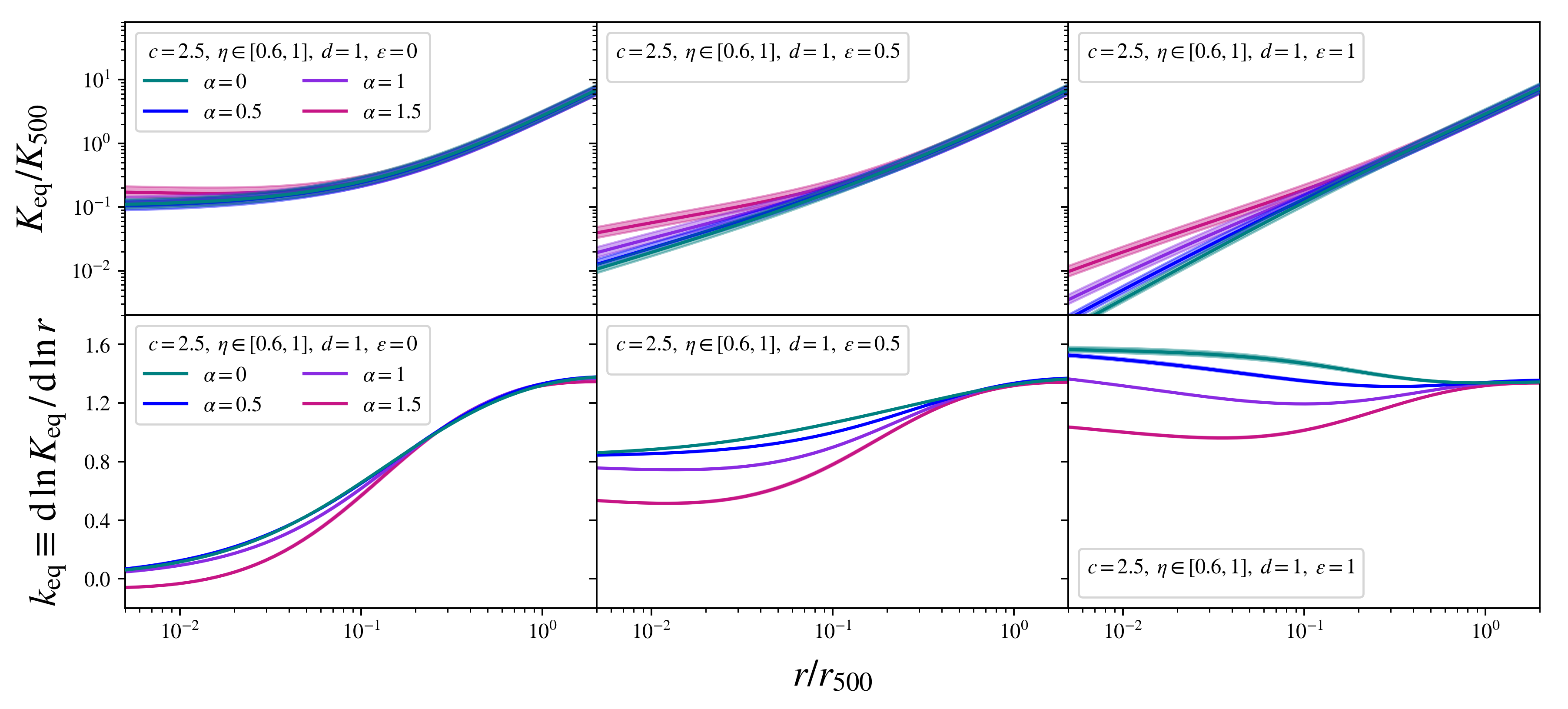}
    \caption{The pristine equilibrium gas entropy profiles, in scale-free form $K_\mathrm{eq}/K_\mathrm{500}$, shown in the top row, and the pristine equilibrium gas entropy slopes, $k_\mathrm{eq} \equiv \mathrm{d} \ln K_\mathrm{eq} / \mathrm{d} \ln r$, shown in the bottom row, each traced over the scaled halocentric radius $r/r_\mathrm{500}$, as predicted for the ideal baryonic cluster halo model. The halo concentration, $c$, and dilution, $d$, are both fixed parameters, whilst each column varies the gas inner slope, $\varepsilon$. Within each box, each colour varies the halo inner slope, $\alpha$, with the solid coloured lines tracing a fraction of cosmological baryon content of $\eta=0.8$, and the shaded colour region around each solid line (not visible for all curves) tracing this value continuously between $\eta=0.6$ and $\eta=1$. }
    \label{Fig - entropy slopes}
\end{figure*}

\section{3. Analysis}
\vspace{2mm}
\subsubsection{The pristine equilibrium gas entropy and gas entropy slope of the ideal baryonic cluster halos}
Taking the expression for the pristine equilibrium gas entropy, Equation \eqref{pristine entropy definition}, we can predict the entropy profiles for the ideal baryonic cluster halo model:
\begin{equation}\label{ideal physical baryonic halo entropy}
\begin{aligned}
    \frac{K_\mathrm{eq}(s, c, \alpha, \eta, d, \varepsilon)}{K_\mathrm{vir}} &= \frac{\left\{3s^\varepsilon \left[1 + \mathcal{C}(c, \alpha, d, \varepsilon)s\right]^{3-\varepsilon}\right\}^{5/3} }{\left[\eta \, \mathcal{U}(c, \alpha, d, \varepsilon)\right]^{2/3}} \\
    & \hspace{-5mm} \times \quad \mathcal{I}(s, c, \alpha, \eta, d, \varepsilon),
\end{aligned}
\end{equation}
as a function of the dimensionless halocentric radius, $s$; the five structural parameters from Table \ref{Parameter table}; and the integral function:
\begin{equation}\label{temperature integral function}
\begin{aligned}
    \mathcal{I}(s, c, \alpha, \eta, d, \varepsilon)&\equiv \int _s ^\infty   \frac{ \mathrm{d}s^\prime \Biggl\{  (1 - \eta f_\mathrm{b, cos})u(c, \alpha) \cdot \int _0 ^{s^\prime} \frac{s^\prime{}^\prime{} ^{2-\alpha} \mathrm{d}s^\prime{}^\prime}{(1 + cs^\prime{}^\prime )^{3-\alpha}}   }{s^\prime{}^{2 + \varepsilon} \left[1 + \mathcal{C}(c, \alpha, d, \varepsilon)s^\prime \right]^{3 - \varepsilon}} \\
    & \hspace{-15mm} + \quad \eta f_\mathrm{b, cos} \mathcal{U}(c, \alpha, d, \varepsilon) \cdot \int _0 ^{s^\prime} \frac{s^\prime{}^\prime{}^{2-\varepsilon} \mathrm{d}s^\prime{}^\prime}{\left[1 + \mathcal{C}(c, \alpha, d, \varepsilon)s^\prime{}^\prime \right]^{3-\varepsilon}} \Biggr\}.
\end{aligned}
\end{equation}
In this expression, the gas entropy is scaled by the virial entropy, $K_\mathrm{vir}$, which we define as:
\begin{equation}\label{virial entropy}
    K_\mathrm{vir} \equiv \left[\frac{\mu_\mathrm{e} m_\mathrm{p}}{f_\mathrm{b, cos} \Delta \rho_\mathrm{crit,0}}\right]^{2/3} k_\mathrm{B}T_\mathrm{vir},
\end{equation}
in terms of the virial temperature,$T_\mathrm{vir}$, defined as:
\begin{equation}
    T_\mathrm{vir} \equiv \frac{1}{3} \frac{\mu m_\mathrm{p}}{k_\mathrm{B}} \frac{GM_\mathrm{vir}}{r_\mathrm{vir}}.
\end{equation}
When $\Delta=500$ in Equation \eqref{virial entropy}, this allows us to define the entropy and temperature values of $K_{500}$ and $T_{500}$. 

Taking the logarithmic derivative of Equation \eqref{ideal physical baryonic halo entropy} with respect to $s$, we find that the pristine equilibrium entropy slope, $k_\mathrm{eq}(s)$, for this model can be solved as:
\begin{equation}\label{ideal physical baryonic halo entropy slope}
\begin{aligned}
    k_\mathrm{eq}(s, c, \alpha, \eta, d, \varepsilon) &= \frac{5}{3} \left[\varepsilon + (3-\varepsilon) \frac{\mathcal{C}(c, \alpha, d, \varepsilon)s}{\left[1 + \mathcal{C}(c, \alpha, d, \varepsilon)s\right]}\right] \\
    &\hspace{-15mm} -  \quad \frac{\Bigl\{ (1 - \eta f_\mathrm{b, cos}) u(c, \alpha) \cdot \int_0 ^s \frac{s^\prime{}^{2-\alpha} \mathrm{d}s^\prime}{(1 + cs^\prime)^{3-\alpha}}  }{\mathcal{I}(s, c, \alpha, \eta, d, \varepsilon) s^{\varepsilon+1}\left[1 + \mathcal{C}(c, \alpha, d, \varepsilon)s\right]^{3-\varepsilon}} \\
    &\hspace{-15mm} + \quad \eta f_\mathrm{b, cos} \mathcal{U}(c, \alpha, d, \varepsilon) \cdot \int_0^s \frac{s^\prime{}^{2-\varepsilon} \mathrm{d}s^\prime}{\left[1 + \mathcal{C}(c, \alpha, d, \varepsilon)s^\prime\right]^{3-\varepsilon}} \Bigr\}.
\end{aligned}
\end{equation}
Over the parameter space detailed in Table \ref{Parameter table}, the profiles for the pristine equilibrium gas entropy, in the form $K_\mathrm{eq}/K_{500}$, and the corresponding slopes, $k_\mathrm{eq}$, are traced within Figure \ref{Fig - entropy slopes}, as a function of the dimensionless halocentric radius $s \equiv r/r_{500}$.

\begin{figure*}[h!]
    \centering
    \includegraphics[width=\textwidth]{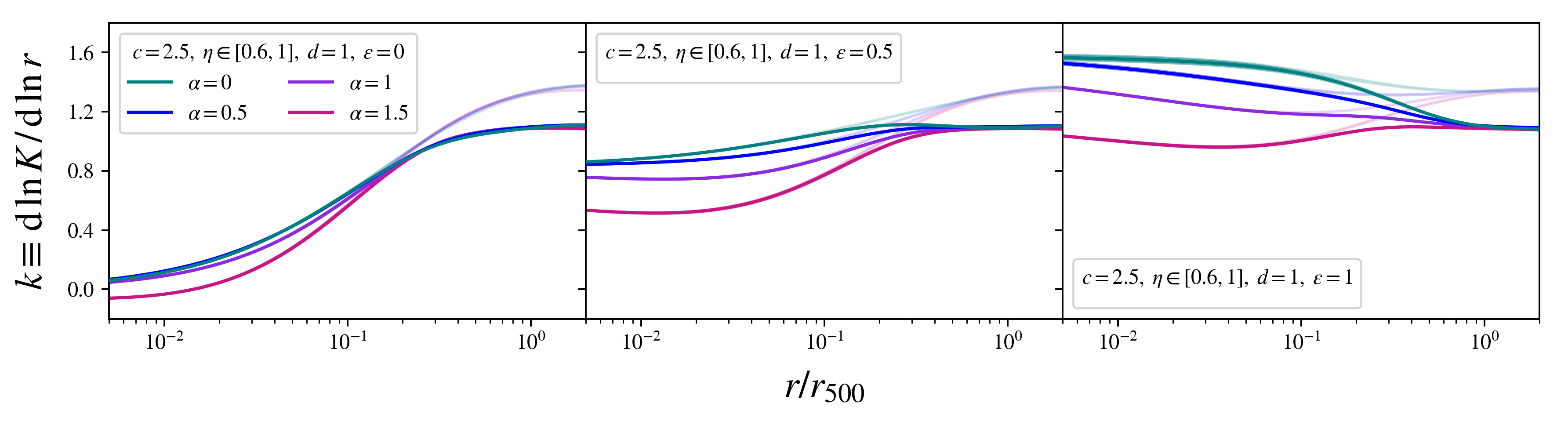}
    \caption{The weighted gas entropy slopes, $k \equiv \mathrm{d} \ln K / \mathrm{d} \ln r$, traced over the scaled halocentric radius $r/r_\mathrm{500}$, derived as a modification to the pristine equilibrium profiles from Figure \ref{Fig - entropy slopes}, when weighted by the weighting function from Equation \eqref{weighting function}. The halo concentration, $c$, and dilution, $d$, are both fixed parameters, whilst each column varies the gas inner slope, $\varepsilon$. Within each box, each colour varies the halo inner slope, $\alpha$, with the solid coloured lines tracing a fraction of cosmological baryon content of $\eta=0.8$, and the shaded colour region around each solid line (not visible for all curves) tracing this value continuously between $\eta=0.6$ and $\eta=1$. The faded profiles in the background of each panel correspond to the associated pristine equilibrium entropy slopes, $k_\mathrm{eq} \equiv \mathrm{d} \ln K_\mathrm{eq} / \mathrm{d} \ln r$, from the top row of Figure \ref{Fig - entropy slopes}.}
    \label{Fig - new entropy slopes}
\end{figure*}

Figure \ref{Fig - entropy slopes} shows how varying the gas profile's inner slope, $\varepsilon$, drives the behaviour of the gas entropy in the central region. Gas cores, $\varepsilon=0$, in the left column, produce high central entropy, characteristic of NCC clusters; weak gas cusps, $\varepsilon=0.5$, in the centre column, attain a central entropy slope of $k_\mathrm{eq} \simeq 0.6-0.9$, which is roughly consistent with observational constraints in CC clusters \autocite[e.g.][]{Babyk2018}. This association between cuspy gas inner slopes and CCs is a well known observational correlation \autocite[see, e.g.][]{Hudson2010}, and is well reproduced in these panels. In the right panel, showing NFW-like gas cusps, $\varepsilon=1$, the gas entropy becomes increasingly steep toward the centre of the cluster, implying a rapid drop in the gas entropy in the core. We note that such steep gradients are not generally observed or predicted. 

Throughout the parameter space traced in Figure \ref{Fig - entropy slopes}, the gas entropy slope converges to a constant value of $k_\mathrm{eq} \simeq 1.4$ in the cluster's outskirts, beyond halocentric radii $r \gtrsim 0.8 r_{500}$. In comparison to the consensus in the literature, where the gas entropy slope is expected to attain a constant value of $k \simeq 1.1$ beyond $r \gtrsim 0.6 r_{500}$, this pristine equilibrium model systematically overestimates the gas entropy in the cluster's outer region. In particular, by Equation \eqref{pristine equilibrium entropy slope}, when treating the intracluster gas density profile as fixed, this overestimate in $k_\mathrm{eq}(s)$ reflects an overestimate in the logarithmic derivative of the pristine equilibrium gas temperature, $T_\mathrm{eq}(s)$, which, as the gas temperature will be decreasing in the cluster's outer region, implies that $T_\mathrm{eq}(s)$ is decreasing too gently with radius in the outskirts. If the gas temperature falls more rapidly, this implies that NTP is required, specifically as an increasing function of halocentric radius, to ensure that the cluster remains in hydrostatic equilibrium.

\subsubsection{Choosing a weighting function}

To relate the pristine and real entropy slopes, and thus predict the required NTP function, we must prescribe the weighting function, $w(s)$. At large radii, when $k_\mathrm{eq} \simeq 1.4$, we require a weighting of $w \simeq 0.8$, such that the entropy slope is reduced to $k \simeq 1.1$, as is observed in both simulations and observations. Leaving the entropy slopes unchanged in the inner region, where slopes consistent with CCs and NCCs are relatively well established, implies that the weighting function must take the form of a continuous step function, transitioning between $w = 1$ and $w = 0.8$ as a function of halocentric radius. 

There are two parameters that need to be chosen for such a function: the steepness of the transition, and the radius at which the transition occurs. We set the mid-point weight of $w = 0.9$ to occur at a halo radius of $r \simeq 0.4 r_{500}$, with the steepness set by an amplitude of $5$ in the exponent. This choice ensures that the entropy slope $k \simeq 1.1$ is reached, and remains fixed, above halocentric radii $r \gtrsim 0.6 r_{500}$, whilst $k \simeq 1$ is a better fit to its value within the region $0.2r_{500} \lesssim r \lesssim 0.4 r_{500}$.

This scale-free weighting function is then specified by the continuous step function:
\begin{equation}\label{weighting function}
    w(s) = 0.8 + \frac{1}{5\left[1 + \mathrm{e}^{5\left[\log_{10}(s) + 0.4\right]}\right]}.
\end{equation}
\textcolor{black}{We emphasise that this choice in parameters within the step function is not unique, and could be altered in both the steepness of the transition and its radial occurrence, both of which exhibit a degree of degeneracy to one another, and each of which can quantitatively impact the predicted NTP profile. However, as we have ensured that our choice produces entropy slopes that are consistent with the values in the literature, we do not consider other choices hereafter. }

These new entropy slopes, calculated by weighting each of the pristine equilibrium entropy slopes, $k_\mathrm{eq}$, from Figure \ref{Fig - entropy slopes}, are shown in Figure \ref{Fig - new entropy slopes}. For all parameter configurations, these weighted entropy slopes now converge to $k = 1.1$ in the cluster's outskirts, as ensured. 

\section{4. Results}

\begin{table*}[h!]
\begin{tabular}{ |p{2.75cm}||p{1.25cm}|p{1.25cm}|p{1.25cm}|}
\hline
 \multicolumn{1}{|c||}{Central NTP fraction} &  \multicolumn{1}{c|}{$A$} & \multicolumn{1}{|c|}{$B$} & \multicolumn{1}{|c|}{$\gamma$} \\ [1ex]
\hline\hline
\rowcolor{white}$\mathcal{F}_0=0$ & $0.501$ & $1.771$ & $1.208$  \\ [2ex]
\hline

\rowcolor{white}$\mathcal{F}_0=0.1$ & $0.451$ & $1.771$ & $1.208$ \\ [2ex]
\hline
\end{tabular}
\caption{Analytic fits for the non-thermal pressure (NTP) fraction, $\mathcal{F} \equiv p_\mathrm{nt} / p$, as a function of the scale-free halocentric radius, $r/r_{500}$, when solved by Equation \eqref{scale-free general solution for the NTP} over the parameter space in Table \ref{Parameter table}, for the weighting function, Equation \eqref{weighting function}, and specified by the cluster's central NTP fraction, $\mathcal{F}_0$, in each choice given below. The best-fitting parameters specify the functional form suggested by \citet{Nelson2014}, given in Equation \eqref{Nelson NTP}.}
\label{Nelson fit table}
\end{table*}

\begin{figure*}[h!]
    \centering
    \includegraphics[width=\textwidth]{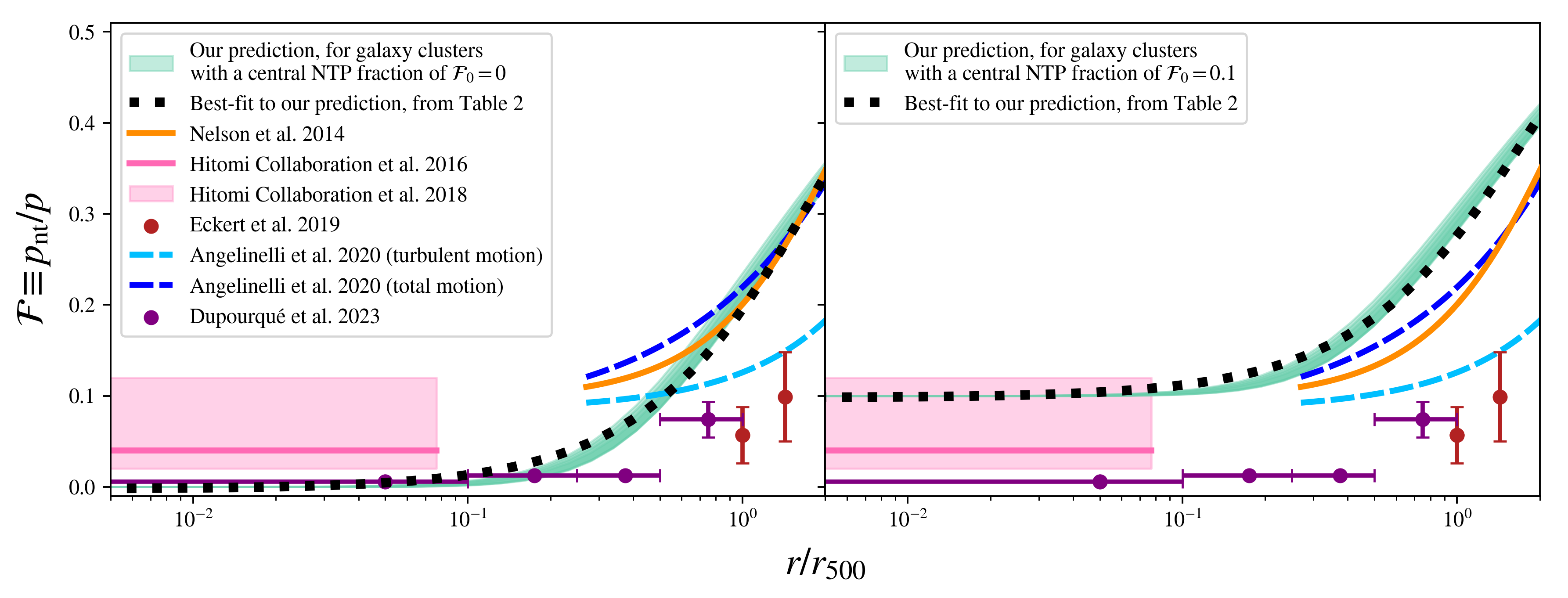}
    \caption{The non-thermal pressure (NTP) fraction, $\mathcal{F} \equiv p_\mathrm{nt} / p$, traced over the scaled halocentric radius $r/r_{500}$, that solves the entropy slope constraints via Equation \eqref{scale-free general solution for the NTP}. In each box, the cluster's structural parameters are varied over the entire parameter space from Table \ref{Parameter table}, producing the turquoise shaded regions, given a choice in the cluster's central NTP fraction, $\mathcal{F}_0$, which is set to $\mathcal{F}_0 = 0$ in the left panel, and $\mathcal{F}_0 = 0.1$ in the right panel. The black dotted line in each box is the best-fit to the functional form proposed in \citet{Nelson2014}, given in Equation \eqref{Nelson NTP}, with its best-fitting parameters specified in Table \ref{Nelson fit table}. We compare our predictions to numerical fits: from \citet{Nelson2014}, shown by the orange line; and from \citet{Angelinelli2020}, shown by the light blue and blue dashed lines, corresponding to different contributions of the gas motion. We also compare to observational constraints: from the \citet{Hitomi2018}, as given by the pink shaded region, with the $4\%$ value from \citet{Hitomi2016} shown by the pink solid line; from \citet{Eckert2019}, shown by the red error bars; and from \citet{Dupourque2023}, shown by the purple error bars.}
    \label{Fig - NTP profiles}
\end{figure*}

\subsubsection{The predicted non-thermal pressure fraction}

\begin{figure*}[h!]
    \centering
    \caption{The gas entropy profiles, in scale-free form $K/K_{500}$, traced over the scaled halocentric radius $r/r_\mathrm{500}$, for the ideal baryonic cluster halos: in pristine equilibrium, in the left panel, indicated by the light blue shaded region; and when including the minimum NTP fraction, as given by the fit for $\mathcal{F}_0=0$ in Table \ref{Nelson fit table}, in the right panel, indicated by the light purple shaded region. These predictions are compared to recent observational fits for the gas entropy profile of galaxy clusters, from \citet{Ghirardini2019}, for samples of cool core clusters (the blue dotted line) and non-cool core clusters (the orange dash-dotted line), as well as to the universal gas entropy profile from \citet{Babyk2018} (the teal dashed line).}
    \includegraphics[width=\textwidth]{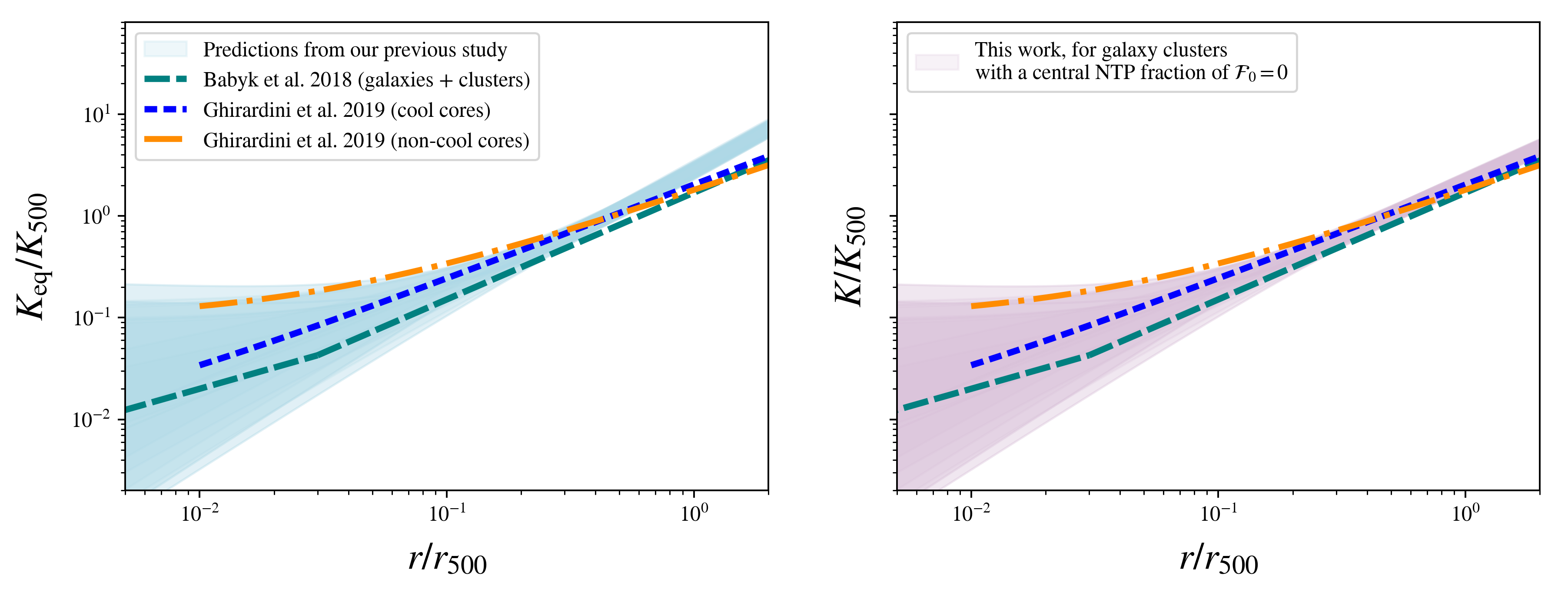}
    \label{Fig - observational comparisons with entropy}
\end{figure*}

We now estimate the scale-free NTP fraction, $\mathcal{F}(s)$, required for the entropy slope of an ideal baryonic cluster halo to be consistent with the imposed constraints, by using 
the weighting function in Equation \eqref{weighting function}. In Figure \ref{Fig - NTP profiles}, we trace these $\mathcal{F}(s)$ profiles over the dimensionless halocentric radius $s\equiv r/r_{500}$, for two choices of the central NTP fraction, $\mathcal{F}_0$, and subsequently evaluated continuously over the parameter space from Table \ref{Parameter table}, producing the turquoise shaded intervals. 

In the left panel, the cluster's central NTP fraction is set to $\mathcal{F}_0 = 0$, commensurate with zero NTP in the cluster's core; this traces the minimum NTP fraction required to attain the imposed entropy constraints. In the right panel, this parameter is set to $\mathcal{F}_0 = 0.1$, corresponding to a baseline NTP fraction of $10\%$ in the cluster's core, as roughly consistent with the \textit{Hitomi} upper limit. Importantly, this central NTP fraction, $\mathcal{F}_0$, does not change the characteristic shape of these NTP fractions; instead, changing this value corresponds to a vertical shift in the NTP fraction over all halocentric radii.  

\textcolor{black}{For each of our two predictions, we fit the parameter space of NTP fraction profiles to the functional form suggested in \citet{Nelson2014}, which is given by the parameterisation:
\begin{equation}\label{Nelson NTP}
    \mathcal{F}(s) = 1 - A \left\{ 1 + \mathrm{e} ^{- (s/B)^\gamma }\right \},
\end{equation}
which we take as a function of the dimensionless halocentric radius $s \equiv r/r_{500}$, consistent with our parameter space in Table \ref{Parameter table}. This fitting procedure allows us to capture the turquoise shaded intervals in Figure \ref{Fig - NTP profiles} with an analytic approximation.} The best-fitting values to the parameters $A$, $B$ and $\gamma$ are given in Table \ref{Nelson fit table}, for each of the two NTP predictions, specified by our two choices in $\mathcal{F}_0$. These best-fit curves are shown by the black dotted lines in each panel of Figure \ref{Fig - NTP profiles}. 

We compare these predictions to the mean profiles obtained in non-radiative hydrodynamic simulations: from \citet{Nelson2014}, shown by the orange line, and from \citet{Angelinelli2020}, \textcolor{black}{shown in both the light blue and blue dashed lines, each predicted from different calculations of the gas motion.} These numerical fits are each given in terms of a mean density radius, $r_{200\textit{m}}$, which we re-scale using the conversion $r_{200\textit{m}} \simeq 2.70 r_{500}$ \autocite[as in, e.g.][]{Nelson2014} to plot in comparison to our model. \textcolor{black}{Further, we show comparison to observational constraints on the NTP: from \citet{Eckert2019}, shown by the red error bars, and from \citet{Dupourque2023}, shown by the purple error bars. We also compare to constraints from the \citet{Hitomi2018} in the cluster's central core, shown by the pink shaded region, with the $4 \% $ value \autocite[as given in][] {Hitomi2016} traced by the pink solid line. }

\textcolor{black}{The left panel of Figure \ref{Fig - NTP profiles} shows that our minimum NTP fraction profile, the $\mathcal{F}_0=0$ result, is in strong agreement with numerical simulation fits at large cluster radii, above $r \gtrsim 0.7 r_{500}$. Whilst this is in strong tension with observational constraints at similar halo radii, this minimum profile is consistent the lower limit of observational constraints available in the cluster's central region \autocite{Hitomi2018, Dupourque2023}.} For this minimum NTP fraction, our model predicts $\mathcal{F} \simeq 0.20$ at $r_{500}$, and $\mathcal{F} \simeq 0.34$ at $2r_{500}$; these predictions are within a few percent of the mean values from \citet{Nelson2014}, and \citet{Angelinelli2020}, in the total gas motion prediction. We use this minimum NTP fraction as our baseline prediction to produce the results that follow below.


\begin{figure*}[h!]
    \centering
    \caption{The gas' temperature and thermal pressure profiles, in scale-free form $T/T_{500}$ and $p_\mathrm{th}/p_{500}$, shown in the top and bottom panels, respectively, each traced over the scaled halocentric radius $r/r_\mathrm{500}$, for the ideal baryonic cluster halos: in pristine equilibrium, in the left panel, indicated by the light blue shaded region; and when including the minimum NTP fraction, as given by the fit for $\mathcal{F}_0=0 $ in Table \ref{Nelson fit table}, in the right panel, indicated by the light purple shaded region. These predictions are compared to recent observational fits for the temperature profile of galaxy clusters, from \citet{Ghirardini2019}, for samples of cool core clusters (the blue dotted line) and non-cool core clusters (the orange dash-dotted line), as well as to the universal gas pressure profile from \citet{Arnaud2010} (the purple dotted line).}
    \includegraphics[width=\textwidth]{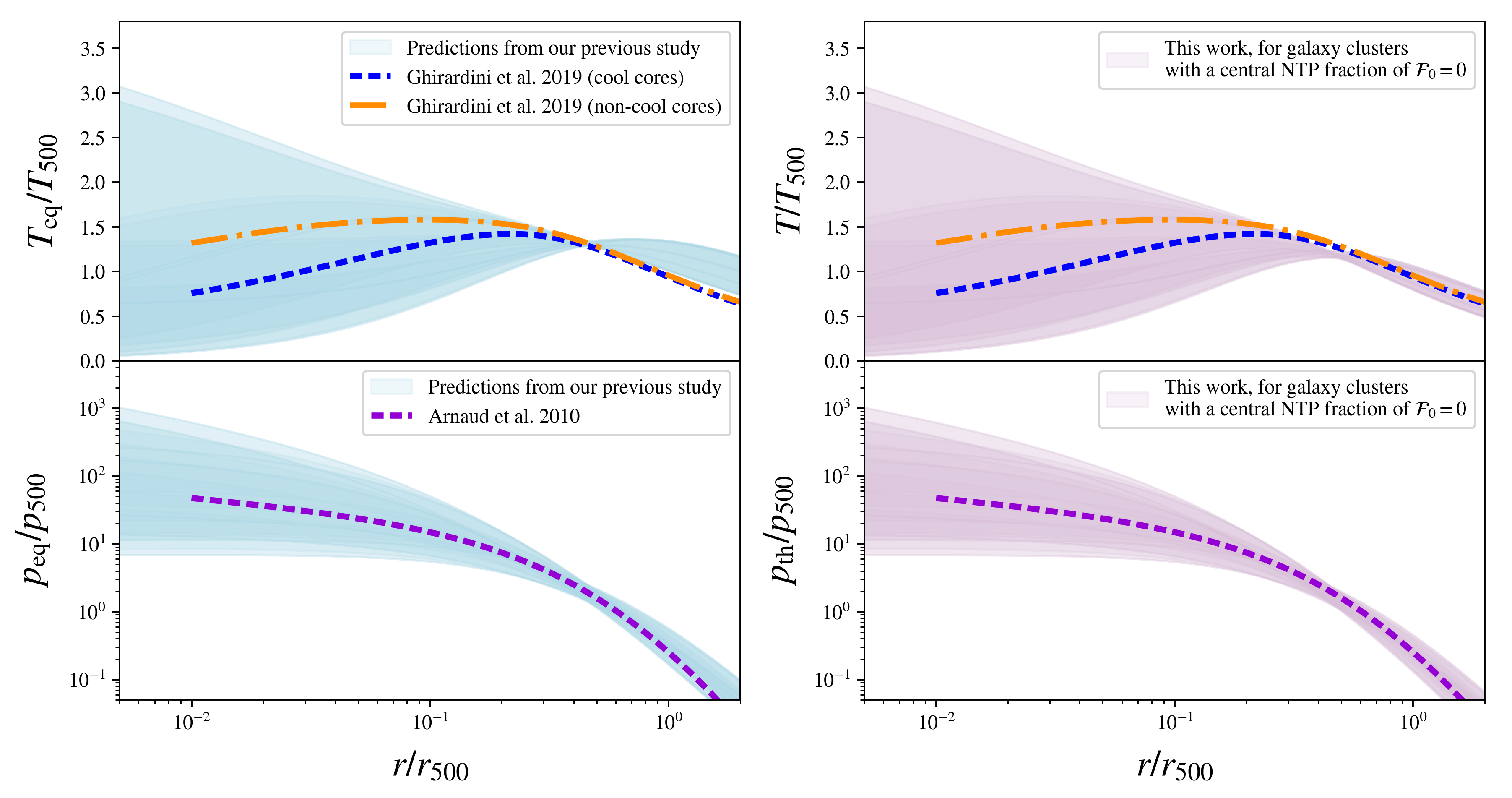}
    \label{Fig - observational comparisons with temperature and pressure}
\end{figure*}

\subsubsection{Implications for the gas entropy, temperature and thermal pressure}

Figure \ref{Fig - observational comparisons with entropy} shows the improvement in predicting the gas' entropy profiles when incorporating our minimum NTP fraction profile in the cluster's hydrostatic state. We show the parameter space of pristine equilibrium entropy profiles, in scale-free form $K_\mathrm{eq}/K_{500}$, in the left panel, shown in the light blue shaded region; this corresponds to the prediction from our previous work. The new parameter space of scale-free entropy profiles, $K/K_{500}$, are given in the right panel, in the light purple shaded region. In each case, the galaxy cluster's structural parameters are those specified in Table \ref{Parameter table}. 

The corresponding scale-free parameter region of gas temperatures, $T/T_{500}$, and thermal pressures, $p_\mathrm{th}/p_{500}$, are traced over this same parameter space,
as shown by the light purple shaded regions in the top right and bottom right panels of Figure \ref{Fig - observational comparisons with temperature and pressure}, respectively.
These profiles are compared to the predictions from our previous work,  S24b, given by the light blue shaded regions in the left panels of this figure: tracing the scale-free parameter region of pristine equilibrium temperatures, $T_\mathrm{eq}/T_{500}$, and equilibrium pressures pressures, $p_\mathrm{eq}/p_{500}$ (as the thermal pressure, without NTP), in the top left and bottom left panels, respectively. Importantly, we see that including the proposed NTP profile in the cluster's hydrostatic state predicts gas temperatures that are now consistent with observational constraints at large cluster radii.

\subsubsection{Implications for the hydrostatic bias}
The NTP fraction at any given halocentric radius of a galaxy cluster will result in a hydrostatic bias, $b(r)$, that arises when estimating the enclosed halo mass, $M(r)$, from its observed thermal properties. This bias is typically quantified \autocite[as in, e.g.][]{Pratt2019, Salvati2019} by the definition:
\begin{equation}\label{hydrostatic bias definition}
    b(r) \equiv 1 - \frac{M_\mathrm{eq}(r)}{M(r)},
\end{equation}
where $M_\mathrm{eq}(r)$ is the halo mass deduced when assuming pristine hydrostatic equilibrium, and $M(r)$ is the halo's real mass. This hydrostatic bias will then be related to the value of the NTP fraction, $\mathcal{F}(r)$, and its first derivative, in the form \autocite[see, e.g.][]{Eckert2019}:
\begin{equation}\label{hydrostatic bias}
    b(r) = \mathcal{F}(r) - \frac{r^2}{\left[1 - \mathcal{F}(r)\right]} \frac{\mathrm{d} \mathcal{F}(r)}{\mathrm{d}r} \frac{p_\mathrm{th}(r)}{GM(r) \rho_\mathrm{gas}(r)}.
\end{equation}
In our scale-free framework, this is equivalent to:
\begin{equation}\label{scale-free hydrostatic bias}
    b(s) = \mathcal{F}(s) - \frac{1}{3}\frac{s^2}{\left[1 - \mathcal{F}(s)\right]} \frac{\mathrm{d} \mathcal{F}(s)}{\mathrm{d}s} \frac{M_\mathrm{vir}}{M(s)}\frac{T(s)}{T_\mathrm{vir}},
\end{equation}
now in terms of the dimensionless halo radius, $s$; a dimensionless ratio of the cluster's true mass, $M(s)$, to its virial mass, $M_\mathrm{vir}$; and a dimensionless ratio of the cluster's temperature, $T(s)$, to its virial temperature, $T_\mathrm{vir}$. By construction, at $r_{500}$, the ratio of the cluster's true mass to the virial mass $M_{500}$ will be unity; similarly, the ratio of the cluster's temperature to the virial temperature $T_{500}$ will be $\simeq 1$ over the chosen parameter space at $r_{500}$ (see, e.g. S24b). The hydrostatic bias can then be estimated by our predicted minimum NTP fraction, which gives $\mathcal{F} \simeq 0.20$ at $r_{500}$, as specified by the best-fit in Table \ref{Nelson fit table}, and the first derivative of this function, with respect to $s$, which will be analytic. 

This NTP profile imposes a hydrostatic bias in the halo mass $M_{500}$ of $b \simeq 0.12$, when measured within $r_{500}$; in other words, when assuming there is no NTP contribution to the hydrostatic state of a galaxy cluster, the halo mass $M_{500}$ will be underestimated by $12\%$ of its real value. This hydrostatic bias will shift the scaling relations of the halo mass $M_{500}$ with respect to the cluster's gas' mean-weighted temperature observables and its integrated SZ signals by approximately this same bias (within $\sim 12 \pm 5\%$ from the results presented in S24b).

\section{5. Conclusion}

We have applied constraints from observed gas entropy slopes to predict the non-thermal pressure (NTP) fraction that is required to explain these observations. 
Our key findings are summarised below.

\begin{itemize}
    \item The required NTP fraction, $\mathcal{F}(r)$, as a function of halocentric radius, $r$, is a radially increasing function, and can be parameterised by its value in the cluster's core, $\mathcal{F}_0$. \\
    
    \item This profile, $\mathcal{F}(r)$, is always well-fit to the functional form proposed in hydrodynamic simulations, as given in \citet{Nelson2014}. \\

    \item The profile for the minimum NTP fraction, defined as the case with $\mathcal{F}_0=0$, is in excellent agreement with the mean NTP fit predicted by numerical simulations, from \citet{Nelson2014} and \citet{Angelinelli2020}, at large halocentric radii, when $r\gtrsim 0.7r_{500}$. \\
    
    \item In the cluster's central region, this minimum NTP fraction is consistent with the lower limit observational constraints from the \citet{Hitomi2018} and \citet{Dupourque2023}, which indicate that clusters have little or no NTP in their core. \\

    \item This profile for the minimum NTP fraction predicts the fractions of $\mathcal{F} \simeq 0.20$ at $r_{500}$, and $\mathcal{F} \simeq 0.34$ at $2r_{500}$. \\
    
    \item Inclusion of this minimum NTP fraction into a hydrostatic equilibrium model predicts entropy, temperature and thermal pressure profiles for the intracluster gas that are consistent with observations. \\

    \item Non-thermal pressure is an important feature in the halo mass scaling relations. Using the minimum NTP fraction results in a hydrostatic bias of $b \simeq 0.12$ when measuring the cluster mass $M_{500}$ within a halocentric radius of $r_{500}$.

\end{itemize}

As noted in the introduction, our expectation is that the NTP profile in a cluster will arise from a combination of gravitationally-driven shocks and mergers, primarily at larger halocentric radii, and feedback processes, such as powerful outflows driven by active galactic nuclei (AGNs), which should manifest at small radius. Our results indicate that the effects of NTP are more pronounced at larger radii, suggesting the important role of gravitational shocks, which are likely to be both strong and long-lived, as predicted by cosmological simulations of clusters \citep[e.g.][]{Power2020}. 

What does this mean for the contribution of feedback to the NTP profile? We observe powerful AGN jets in galaxy clusters \citep[e.g.][]{Shabala2018} but they are understood to be intermittent, both in their observable properties and in the manner in which they impact their environment \citep[e.g.][]{Yates2018}. If we assume that feedback will be driven by jet dynamics and energetics, what does the implied form of the NTP profile at small radii mean for our physical understanding of the action of feedback? We will investigate this question in a forthcoming paper that investigates the NTP in the central region of galaxy clusters, specifically, whether or not realistic AGN feedback is consistent with zero NTP in the central regions of galaxy clusters. 

\vspace{1cm}


 AS acknowledges the support of the Australian Government Research Training Program Fees Offset; the Bruce and Betty Green Postgraduate Research Scholarship; and The University Club of Western Australia Research Travel Scholarship. AS and CP acknowledge the support of the ARC Centre of Excellence for All Sky Astrophysics in 3 Dimensions (ASTRO 3D), through project number CE170100013. CB gratefully acknowledges support from the Forrest Research Foundation.

\printbibliography

@ARTICLE{Planck2016,
       author = {{Planck Collaboration} and {Ade}, P.~A.~R. and {Aghanim}, N. and {Arnaud}, M. and {Ashdown}, M. and {Aumont}, J. and {Baccigalupi}, C. and {Banday}, A.~J. and {Barreiro}, R.~B. and {Bartlett}, J.~G. and {Bartolo}, N. and {Battaner}, E. and {Battye}, R. and {Benabed}, K. and {Beno{\^\i}t}, A. and {Benoit-L{\'e}vy}, A. and {Bernard}, J. -P. and {Bersanelli}, M. and {Bielewicz}, P. and {Bock}, J.~J. and {Bonaldi}, A. and {Bonavera}, L. and {Bond}, J.~R. and {Borrill}, J. and {Bouchet}, F.~R. and {Boulanger}, F. and {Bucher}, M. and {Burigana}, C. and {Butler}, R.~C. and {Calabrese}, E. and {Cardoso}, J. -F. and {Catalano}, A. and {Challinor}, A. and {Chamballu}, A. and {Chary}, R. -R. and {Chiang}, H.~C. and {Chluba}, J. and {Christensen}, P.~R. and {Church}, S. and {Clements}, D.~L. and {Colombi}, S. and {Colombo}, L.~P.~L. and {Combet}, C. and {Coulais}, A. and {Crill}, B.~P. and {Curto}, A. and {Cuttaia}, F. and {Danese}, L. and {Davies}, R.~D. and {Davis}, R.~J. and {de Bernardis}, P. and {de Rosa}, A. and {de Zotti}, G. and {Delabrouille}, J. and {D{\'e}sert}, F. -X. and {Di Valentino}, E. and {Dickinson}, C. and {Diego}, J.~M. and {Dolag}, K. and {Dole}, H. and {Donzelli}, S. and {Dor{\'e}}, O. and {Douspis}, M. and {Ducout}, A. and {Dunkley}, J. and {Dupac}, X. and {Efstathiou}, G. and {Elsner}, F. and {En{\ss}lin}, T.~A. and {Eriksen}, H.~K. and {Farhang}, M. and {Fergusson}, J. and {Finelli}, F. and {Forni}, O. and {Frailis}, M. and {Fraisse}, A.~A. and {Franceschi}, E. and {Frejsel}, A. and {Galeotta}, S. and {Galli}, S. and {Ganga}, K. and {Gauthier}, C. and {Gerbino}, M. and {Ghosh}, T. and {Giard}, M. and {Giraud-H{\'e}raud}, Y. and {Giusarma}, E. and {Gjerl{\o}w}, E. and {Gonz{\'a}lez-Nuevo}, J. and {G{\'o}rski}, K.~M. and {Gratton}, S. and {Gregorio}, A. and {Gruppuso}, A. and {Gudmundsson}, J.~E. and {Hamann}, J. and {Hansen}, F.~K. and {Hanson}, D. and {Harrison}, D.~L. and {Helou}, G. and {Henrot-Versill{\'e}}, S. and {Hern{\'a}ndez-Monteagudo}, C. and {Herranz}, D. and {Hildebrandt}, S.~R. and {Hivon}, E. and {Hobson}, M. and {Holmes}, W.~A. and {Hornstrup}, A. and {Hovest}, W. and {Huang}, Z. and {Huffenberger}, K.~M. and {Hurier}, G. and {Jaffe}, A.~H. and {Jaffe}, T.~R. and {Jones}, W.~C. and {Juvela}, M. and {Keih{\"a}nen}, E. and {Keskitalo}, R. and {Kisner}, T.~S. and {Kneissl}, R. and {Knoche}, J. and {Knox}, L. and {Kunz}, M. and {Kurki-Suonio}, H. and {Lagache}, G. and {L{\"a}hteenm{\"a}ki}, A. and {Lamarre}, J. -M. and {Lasenby}, A. and {Lattanzi}, M. and {Lawrence}, C.~R. and {Leahy}, J.~P. and {Leonardi}, R. and {Lesgourgues}, J. and {Levrier}, F. and {Lewis}, A. and {Liguori}, M. and {Lilje}, P.~B. and {Linden-V{\o}rnle}, M. and {L{\'o}pez-Caniego}, M. and {Lubin}, P.~M. and {Mac{\'\i}as-P{\'e}rez}, J.~F. and {Maggio}, G. and {Maino}, D. and {Mandolesi}, N. and {Mangilli}, A. and {Marchini}, A. and {Maris}, M. and {Martin}, P.~G. and {Martinelli}, M. and {Mart{\'\i}nez-Gonz{\'a}lez}, E. and {Masi}, S. and {Matarrese}, S. and {McGehee}, P. and {Meinhold}, P.~R. and {Melchiorri}, A. and {Melin}, J. -B. and {Mendes}, L. and {Mennella}, A. and {Migliaccio}, M. and {Millea}, M. and {Mitra}, S. and {Miville-Desch{\^e}nes}, M. -A. and {Moneti}, A. and {Montier}, L. and {Morgante}, G. and {Mortlock}, D. and {Moss}, A. and {Munshi}, D. and {Murphy}, J.~A. and {Naselsky}, P. and {Nati}, F. and {Natoli}, P. and {Netterfield}, C.~B. and {N{\o}rgaard-Nielsen}, H.~U. and {Noviello}, F. and {Novikov}, D. and {Novikov}, I. and {Oxborrow}, C.~A. and {Paci}, F. and {Pagano}, L. and {Pajot}, F. and {Paladini}, R. and {Paoletti}, D. and {Partridge}, B. and {Pasian}, F. and {Patanchon}, G. and {Pearson}, T.~J. and {Perdereau}, O. and {Perotto}, L. and {Perrotta}, F. and {Pettorino}, V. and {Piacentini}, F. and {Piat}, M. and {Pierpaoli}, E. and {Pietrobon}, D. and {Plaszczynski}, S. and {Pointecouteau}, E. and {Polenta}, G. and {Popa}, L. and {Pratt}, G.~W. and {Pr{\'e}zeau}, G. and {Prunet}, S. and {Puget}, J. -L. and {Rachen}, J.~P. and {Reach}, W.~T. and {Rebolo}, R. and {Reinecke}, M. and {Remazeilles}, M. and {Renault}, C. and {Renzi}, A. and {Ristorcelli}, I. and {Rocha}, G. and {Rosset}, C. and {Rossetti}, M. and {Roudier}, G. and {Rouill{\'e} d'Orfeuil}, B. and {Rowan-Robinson}, M. and {Rubi{\~n}o-Mart{\'\i}n}, J.~A. and {Rusholme}, B. and {Said}, N. and {Salvatelli}, V. and {Salvati}, L. and {Sandri}, M. and {Santos}, D. and {Savelainen}, M. and {Savini}, G. and {Scott}, D. and {Seiffert}, M.~D. and {Serra}, P. and {Shellard}, E.~P.~S. and {Spencer}, L.~D. and {Spinelli}, M. and {Stolyarov}, V. and {Stompor}, R. and {Sudiwala}, R. and {Sunyaev}, R. and {Sutton}, D. and {Suur-Uski}, A. -S. and {Sygnet}, J. -F. and {Tauber}, J.~A. and {Terenzi}, L. and {Toffolatti}, L. and {Tomasi}, M. and {Tristram}, M. and {Trombetti}, T. and {Tucci}, M. and {Tuovinen}, J. and {T{\"u}rler}, M. and {Umana}, G. and {Valenziano}, L. and {Valiviita}, J. and {Van Tent}, F. and {Vielva}, P. and {Villa}, F. and {Wade}, L.~A. and {Wandelt}, B.~D. and {Wehus}, I.~K. and {White}, M. and {White}, S.~D.~M. and {Wilkinson}, A. and {Yvon}, D. and {Zacchei}, A. and {Zonca}, A.},
        title = "{Planck 2015 results. XIII. Cosmological parameters}",
      journal = {\aap},
         year = 2016,
        month = sep,
       volume = {594},
          eid = {A13},
        pages = {A13},
          doi = {10.1051/0004-6361/201525830},
archivePrefix = {arXiv},
       eprint = {1502.01589},
 primaryClass = {astro-ph.CO},
       adsurl = {https://ui.adsabs.harvard.edu/abs/2016A&A...594A..13P}
}

@ARTICLE{Babyk2018,
       author = {{Babyk}, Iu. V. and {McNamara}, B.~R. and {Nulsen}, P.~E.~J. and {Russell}, H.~R. and {Vantyghem}, A.~N. and {Hogan}, M.~T. and {Pulido}, F.~A.},
        title = "{A Universal Entropy Profile for the Hot Atmospheres of Galaxies and Clusters within R $_{2500}$}",
      journal = {\apj},
         year = 2018,
        month = jul,
       volume = {862},
       number = {1},
          eid = {39},
        pages = {39},
          doi = {10.3847/1538-4357/aacce5},
archivePrefix = {arXiv},
       eprint = {1802.02589},
 primaryClass = {astro-ph.CO},
       adsurl = {https://ui.adsabs.harvard.edu/abs/2018ApJ...862...39B}
}

@ARTICLE{Nelson2014,
       author = {{Nelson}, Kaylea and {Lau}, Erwin T. and {Nagai}, Daisuke},
        title = "{Hydrodynamic Simulation of Non-thermal Pressure Profiles of Galaxy Clusters}",
      journal = {\apj},
         year = 2014,
        month = sep,
       volume = {792},
       number = {1},
          eid = {25},
        pages = {25},
          doi = {10.1088/0004-637X/792/1/25},
archivePrefix = {arXiv},
       eprint = {1404.4636},
 primaryClass = {astro-ph.CO},
       adsurl = {https://ui.adsabs.harvard.edu/abs/2014ApJ...792...25N}
}

@ARTICLE{Sayers2021,
       author = {{Sayers}, Jack and {Sereno}, Mauro and {Ettori}, Stefano and {Rasia}, Elena and {Cui}, Weiguang and {Golwala}, Sunil and {Umetsu}, Keiichi and {Yepes}, Gustavo},
        title = "{CLUMP-3D: the lack of non-thermal motions in galaxy cluster cores}",
      journal = {\mnras},
         year = 2021,
        month = aug,
       volume = {505},
       number = {3},
        pages = {4338-4344},
          doi = {10.1093/mnras/stab1542},
archivePrefix = {arXiv},
       eprint = {2102.06324},
 primaryClass = {astro-ph.CO},
       adsurl = {https://ui.adsabs.harvard.edu/abs/2021MNRAS.505.4338S}
}

@ARTICLE{Eckert2019,
       author = {{Eckert}, D. and {Ghirardini}, V. and {Ettori}, S. and {Rasia}, E. and {Biffi}, V. and {Pointecouteau}, E. and {Rossetti}, M. and {Molendi}, S. and {Vazza}, F. and {Gastaldello}, F. and {Gaspari}, M. and {De Grandi}, S. and {Ghizzardi}, S. and {Bourdin}, H. and {Tchernin}, C. and {Roncarelli}, M.},
        title = "{Non-thermal pressure support in X-COP galaxy clusters}",
      journal = {\aap},
         year = 2019,
        month = jan,
       volume = {621},
          eid = {A40},
        pages = {A40},
          doi = {10.1051/0004-6361/201833324},
archivePrefix = {arXiv},
       eprint = {1805.00034},
 primaryClass = {astro-ph.CO},
       adsurl = {https://ui.adsabs.harvard.edu/abs/2019A&A...621A..40E}
}

@ARTICLE{Siegel2018,
       author = {{Siegel}, Seth R. and {Sayers}, Jack and {Mahdavi}, Andisheh and {Donahue}, Megan and {Merten}, Julian and {Zitrin}, Adi and {Meneghetti}, Massimo and {Umetsu}, Keiichi and {Czakon}, Nicole G. and {Golwala}, Sunil R. and {Postman}, Marc and {Koch}, Patrick M. and {Koekemoer}, Anton M. and {Lin}, Kai-Yang and {Melchior}, Peter and {Molnar}, Sandor M. and {Moustakas}, Leonidas and {Mroczkowski}, Tony K. and {Pierpaoli}, Elena and {Shitanishi}, Jennifer},
        title = "{Constraints on the Mass, Concentration, and Nonthermal Pressure Support of Six CLASH Clusters from a Joint Analysis of X-Ray, SZ, and Lensing Data}",
      journal = {\apj},
         year = 2018,
        month = jul,
       volume = {861},
       number = {1},
          eid = {71},
        pages = {71},
          doi = {10.3847/1538-4357/aac5f8},
archivePrefix = {arXiv},
       eprint = {1612.05377},
 primaryClass = {astro-ph.CO},
       adsurl = {https://ui.adsabs.harvard.edu/abs/2018ApJ...861...71S}
}

@ARTICLE{Martizzi2016,
       author = {{Martizzi}, Davide and {Agrusa}, Harrison},
        title = "{Mass modeling of galaxy clusters: quantifying hydrostatic bias and contribution from non-thermal pressure}",
      journal = {arXiv e-prints},
         year = 2016,
        month = aug,
          eid = {arXiv:1608.04388},
        pages = {arXiv:1608.04388},
          doi = {10.48550/arXiv.1608.04388},
archivePrefix = {arXiv},
       eprint = {1608.04388},
 primaryClass = {astro-ph.CO},
       adsurl = {https://ui.adsabs.harvard.edu/abs/2016arXiv160804388M}
}

@ARTICLE{Angelinelli2020,
       author = {{Angelinelli}, M. and {Vazza}, F. and {Giocoli}, C. and {Ettori}, S. and {Jones}, T.~W. and {Brunetti}, G. and {Br{\"u}ggen}, M. and {Eckert}, D.},
        title = "{Turbulent pressure support and hydrostatic mass bias in the intracluster medium}",
      journal = {\mnras},
         year = 2020,
        month = jun,
       volume = {495},
       number = {1},
        pages = {864-885},
          doi = {10.1093/mnras/staa975},
archivePrefix = {arXiv},
       eprint = {1905.04896},
 primaryClass = {astro-ph.CO},
       adsurl = {https://ui.adsabs.harvard.edu/abs/2020MNRAS.495..864A}
}

@ARTICLE{Pearce2020,
       author = {{Pearce}, Francesca A. and {Kay}, Scott T. and {Barnes}, David J. and {Bower}, Richard G. and {Schaller}, Matthieu},
        title = "{Hydrostatic mass estimates of massive galaxy clusters: a study with varying hydrodynamics flavours and non-thermal pressure support}",
      journal = {\mnras},
         year = 2020,
        month = jan,
       volume = {491},
       number = {2},
        pages = {1622-1642},
          doi = {10.1093/mnras/stz3003},
archivePrefix = {arXiv},
       eprint = {1910.10217},
 primaryClass = {astro-ph.CO},
       adsurl = {https://ui.adsabs.harvard.edu/abs/2020MNRAS.491.1622P}
}

@ARTICLE{Vikhlinin2006,
       author = {{Vikhlinin}, A. and {Kravtsov}, A. and {Forman}, W. and {Jones}, C. and {Markevitch}, M. and {Murray}, S.~S. and {Van Speybroeck}, L.},
        title = "{Chandra Sample of Nearby Relaxed Galaxy Clusters: Mass, Gas Fraction, and Mass-Temperature Relation}",
      journal = {\apj},
         year = 2006,
        month = apr,
       volume = {640},
       number = {2},
        pages = {691-709},
          doi = {10.1086/500288},
archivePrefix = {arXiv},
       eprint = {astro-ph/0507092},
 primaryClass = {astro-ph},
       adsurl = {https://ui.adsabs.harvard.edu/abs/2006ApJ...640..691V}
}

@ARTICLE{Vikhlinin2009,
       author = {{Vikhlinin}, A. and {Burenin}, R.~A. and {Ebeling}, H. and {Forman}, W.~R. and {Hornstrup}, A. and {Jones}, C. and {Kravtsov}, A.~V. and {Murray}, S.~S. and {Nagai}, D. and {Quintana}, H. and {Voevodkin}, A.},
        title = "{Chandra Cluster Cosmology Project. II. Samples and X-Ray Data Reduction}",
      journal = {\apj},
         year = 2009,
        month = feb,
       volume = {692},
       number = {2},
        pages = {1033-1059},
          doi = {10.1088/0004-637X/692/2/1033},
archivePrefix = {arXiv},
       eprint = {0805.2207},
 primaryClass = {astro-ph},
       adsurl = {https://ui.adsabs.harvard.edu/abs/2009ApJ...692.1033V}
}

@ARTICLE{Babyk2023,
       author = {{Babyk}, Iurii V. and {McNamara}, Brian R.},
        title = "{The Halo Mass-Temperature Relation for Clusters, Groups, and Galaxies}",
      journal = {\apj},
         year = 2023,
        month = mar,
       volume = {946},
       number = {1},
          eid = {54},
        pages = {54},
          doi = {10.3847/1538-4357/acbf4b},
archivePrefix = {arXiv},
       eprint = {2302.11247},
 primaryClass = {astro-ph.GA},
       adsurl = {https://ui.adsabs.harvard.edu/abs/2023ApJ...946...54B}
}

@ARTICLE{Vanderlinde2010,
       author = {{Vanderlinde}, K. and {Crawford}, T.~M. and {de Haan}, T. and {Dudley}, J.~P. and {Shaw}, L. and {Ade}, P.~A.~R. and {Aird}, K.~A. and {Benson}, B.~A. and {Bleem}, L.~E. and {Brodwin}, M. and {Carlstrom}, J.~E. and {Chang}, C.~L. and {Crites}, A.~T. and {Desai}, S. and {Dobbs}, M.~A. and {Foley}, R.~J. and {George}, E.~M. and {Gladders}, M.~D. and {Hall}, N.~R. and {Halverson}, N.~W. and {High}, F.~W. and {Holder}, G.~P. and {Holzapfel}, W.~L. and {Hrubes}, J.~D. and {Joy}, M. and {Keisler}, R. and {Knox}, L. and {Lee}, A.~T. and {Leitch}, E.~M. and {Loehr}, A. and {Lueker}, M. and {Marrone}, D.~P. and {McMahon}, J.~J. and {Mehl}, J. and {Meyer}, S.~S. and {Mohr}, J.~J. and {Montroy}, T.~E. and {Ngeow}, C. -C. and {Padin}, S. and {Plagge}, T. and {Pryke}, C. and {Reichardt}, C.~L. and {Rest}, A. and {Ruel}, J. and {Ruhl}, J.~E. and {Schaffer}, K.~K. and {Shirokoff}, E. and {Song}, J. and {Spieler}, H.~G. and {Stalder}, B. and {Staniszewski}, Z. and {Stark}, A.~A. and {Stubbs}, C.~W. and {van Engelen}, A. and {Vieira}, J.~D. and {Williamson}, R. and {Yang}, Y. and {Zahn}, O. and {Zenteno}, A.},
        title = "{Galaxy Clusters Selected with the Sunyaev-Zel'dovich Effect from 2008 South Pole Telescope Observations}",
      journal = {\apj},
         year = 2010,
        month = oct,
       volume = {722},
       number = {2},
        pages = {1180-1196},
          doi = {10.1088/0004-637X/722/2/1180},
archivePrefix = {arXiv},
       eprint = {1003.0003},
 primaryClass = {astro-ph.CO},
       adsurl = {https://ui.adsabs.harvard.edu/abs/2010ApJ...722.1180V}
}

@ARTICLE{Andersson2011,
       author = {{Andersson}, K. and {Benson}, B.~A. and {Ade}, P.~A.~R. and {Aird}, K.~A. and {Armstrong}, B. and {Bautz}, M. and {Bleem}, L.~E. and {Brodwin}, M. and {Carlstrom}, J.~E. and {Chang}, C.~L. and {Crawford}, T.~M. and {Crites}, A.~T. and {de Haan}, T. and {Desai}, S. and {Dobbs}, M.~A. and {Dudley}, J.~P. and {Foley}, R.~J. and {Forman}, W.~R. and {Garmire}, G. and {George}, E.~M. and {Gladders}, M.~D. and {Halverson}, N.~W. and {High}, F.~W. and {Holder}, G.~P. and {Holzapfel}, W.~L. and {Hrubes}, J.~D. and {Jones}, C. and {Joy}, M. and {Keisler}, R. and {Knox}, L. and {Lee}, A.~T. and {Leitch}, E.~M. and {Lueker}, M. and {Marrone}, D.~P. and {McMahon}, J.~J. and {Mehl}, J. and {Meyer}, S.~S. and {Mohr}, J.~J. and {Montroy}, T.~E. and {Murray}, S.~S. and {Padin}, S. and {Plagge}, T. and {Pryke}, C. and {Reichardt}, C.~L. and {Rest}, A. and {Ruel}, J. and {Ruhl}, J.~E. and {Schaffer}, K.~K. and {Shaw}, L. and {Shirokoff}, E. and {Song}, J. and {Spieler}, H.~G. and {Stalder}, B. and {Staniszewski}, Z. and {Stark}, A.~A. and {Stubbs}, C.~W. and {Vanderlinde}, K. and {Vieira}, J.~D. and {Vikhlinin}, A. and {Williamson}, R. and {Yang}, Y. and {Zahn}, O. and {Zenteno}, A.},
        title = "{X-Ray Properties of the First Sunyaev-Zel'dovich Effect Selected Galaxy Cluster Sample from the South Pole Telescope}",
      journal = {\apj},
         year = 2011,
        month = sep,
       volume = {738},
       number = {1},
          eid = {48},
        pages = {48},
          doi = {10.1088/0004-637X/738/1/48},
archivePrefix = {arXiv},
       eprint = {1006.3068},
 primaryClass = {astro-ph.CO},
       adsurl = {https://ui.adsabs.harvard.edu/abs/2011ApJ...738...48A}
}

@ARTICLE{SZ1972,
       author = {{Sunyaev}, R.~A. and {Zeldovich}, Ya. B.},
        title = "{The Observations of Relic Radiation as a Test of the Nature of X-Ray Radiation from the Clusters of Galaxies}",
      journal = {Comments on Astrophysics and Space Physics},
         year = 1972,
        month = nov,
       volume = {4},
        pages = {173},
       adsurl = {https://ui.adsabs.harvard.edu/abs/1972CoASP...4..173S}
}

@ARTICLE{SZ1970,
       author = {{Sunyaev}, R.~A. and {Zeldovich}, Ya. B.},
        title = "{The Spectrum of Primordial Radiation, its Distortions and their Significance}",
      journal = {Comments on Astrophysics and Space Physics},
         year = 1970,
        month = mar,
       volume = {2},
        pages = {66},
       adsurl = {https://ui.adsabs.harvard.edu/abs/1970CoASP...2...66S}
}

@INPROCEEDINGS{Chandra,
       author = {{Weisskopf}, Martin C. and {Tananbaum}, Harvey D. and {Van Speybroeck}, Leon P. and {O'Dell}, Stephen L.},
        title = "{Chandra X-ray Observatory (CXO): overview}",
    booktitle = {X-Ray Optics, Instruments, and Missions III},
         year = 2000,
       editor = {{Truemper}, Joachim E. and {Aschenbach}, Bernd},
       series = {Society of Photo-Optical Instrumentation Engineers (SPIE) Conference Series},
       volume = {4012},
        month = jul,
        pages = {2-16},
          doi = {10.1117/12.391545},
archivePrefix = {arXiv},
       eprint = {astro-ph/0004127},
 primaryClass = {astro-ph},
       adsurl = {https://ui.adsabs.harvard.edu/abs/2000SPIE.4012....2W}
}

@ARTICLE{XMMNewton,
       author = {{Jansen}, F. and {Lumb}, D. and {Altieri}, B. and {Clavel}, J. and {Ehle}, M. and {Erd}, C. and {Gabriel}, C. and {Guainazzi}, M. and {Gondoin}, P. and {Much}, R. and {Munoz}, R. and {Santos}, M. and {Schartel}, N. and {Texier}, D. and {Vacanti}, G.},
        title = "{XMM-Newton observatory. I. The spacecraft and operations}",
      journal = {\aap},
         year = 2001,
        month = jan,
       volume = {365},
        pages = {L1-L6},
          doi = {10.1051/0004-6361:20000036},
       adsurl = {https://ui.adsabs.harvard.edu/abs/2001A&A...365L...1J}
}

@ARTICLE{eROSITA,
       author = {{Predehl}, P. and {Andritschke}, R. and {Arefiev}, V. and {Babyshkin}, V. and {Batanov}, O. and {Becker}, W. and {B{\"o}hringer}, H. and {Bogomolov}, A. and {Boller}, T. and {Borm}, K. and {Bornemann}, W. and {Br{\"a}uninger}, H. and {Br{\"u}ggen}, M. and {Brunner}, H. and {Brusa}, M. and {Bulbul}, E. and {Buntov}, M. and {Burwitz}, V. and {Burkert}, W. and {Clerc}, N. and {Churazov}, E. and {Coutinho}, D. and {Dauser}, T. and {Dennerl}, K. and {Doroshenko}, V. and {Eder}, J. and {Emberger}, V. and {Eraerds}, T. and {Finoguenov}, A. and {Freyberg}, M. and {Friedrich}, P. and {Friedrich}, S. and {F{\"u}rmetz}, M. and {Georgakakis}, A. and {Gilfanov}, M. and {Granato}, S. and {Grossberger}, C. and {Gueguen}, A. and {Gureev}, P. and {Haberl}, F. and {H{\"a}lker}, O. and {Hartner}, G. and {Hasinger}, G. and {Huber}, H. and {Ji}, L. and {Kienlin}, A. v. and {Kink}, W. and {Korotkov}, F. and {Kreykenbohm}, I. and {Lamer}, G. and {Lomakin}, I. and {Lapshov}, I. and {Liu}, T. and {Maitra}, C. and {Meidinger}, N. and {Menz}, B. and {Merloni}, A. and {Mernik}, T. and {Mican}, B. and {Mohr}, J. and {M{\"u}ller}, S. and {Nandra}, K. and {Nazarov}, V. and {Pacaud}, F. and {Pavlinsky}, M. and {Perinati}, E. and {Pfeffermann}, E. and {Pietschner}, D. and {Ramos-Ceja}, M.~E. and {Rau}, A. and {Reiffers}, J. and {Reiprich}, T.~H. and {Robrade}, J. and {Salvato}, M. and {Sanders}, J. and {Santangelo}, A. and {Sasaki}, M. and {Scheuerle}, H. and {Schmid}, C. and {Schmitt}, J. and {Schwope}, A. and {Shirshakov}, A. and {Steinmetz}, M. and {Stewart}, I. and {Str{\"u}der}, L. and {Sunyaev}, R. and {Tenzer}, C. and {Tiedemann}, L. and {Tr{\"u}mper}, J. and {Voron}, V. and {Weber}, P. and {Wilms}, J. and {Yaroshenko}, V.},
        title = "{The eROSITA X-ray telescope on SRG}",
      journal = {\aap},
         year = 2021,
        month = mar,
       volume = {647},
          eid = {A1},
        pages = {A1},
          doi = {10.1051/0004-6361/202039313},
archivePrefix = {arXiv},
       eprint = {2010.03477},
 primaryClass = {astro-ph.HE},
       adsurl = {https://ui.adsabs.harvard.edu/abs/2021A&A...647A...1P}
}

@ARTICLE{Ettori2022,
       author = {{Ettori}, S. and {Eckert}, D.},
        title = "{Tracing the non-thermal pressure and hydrostatic bias in galaxy clusters}",
      journal = {\aap},
         year = 2022,
        month = jan,
       volume = {657},
          eid = {L1},
        pages = {L1},
          doi = {10.1051/0004-6361/202142638},
archivePrefix = {arXiv},
       eprint = {2112.07554},
 primaryClass = {astro-ph.CO},
       adsurl = {https://ui.adsabs.harvard.edu/abs/2022A&A...657L...1E}
}

@ARTICLE{Voit2005,
       author = {{Voit}, G. Mark and {Kay}, Scott T. and {Bryan}, Greg L.},
        title = "{The baseline intracluster entropy profile from gravitational structure formation}",
      journal = {\mnras},
         year = 2005,
        month = dec,
       volume = {364},
       number = {3},
        pages = {909-916},
          doi = {10.1111/j.1365-2966.2005.09621.x},
archivePrefix = {arXiv},
       eprint = {astro-ph/0511252},
 primaryClass = {astro-ph},
       adsurl = {https://ui.adsabs.harvard.edu/abs/2005MNRAS.364..909V}
}

@ARTICLE{Hogan2017,
       author = {{Hogan}, M.~T. and {McNamara}, B.~R. and {Pulido}, F. and {Nulsen}, P.~E.~J. and {Russell}, H.~R. and {Vantyghem}, A.~N. and {Edge}, A.~C. and {Main}, R.~A.},
        title = "{Mass Distribution in Galaxy Cluster Cores}",
      journal = {\apj},
         year = 2017,
        month = mar,
       volume = {837},
       number = {1},
          eid = {51},
        pages = {51},
          doi = {10.3847/1538-4357/aa5f56},
archivePrefix = {arXiv},
       eprint = {1610.04617},
 primaryClass = {astro-ph.GA},
       adsurl = {https://ui.adsabs.harvard.edu/abs/2017ApJ...837...51H}
}

@ARTICLE{Hudson2010,
       author = {{Hudson}, D.~S. and {Mittal}, R. and {Reiprich}, T.~H. and {Nulsen}, P.~E.~J. and {Andernach}, H. and {Sarazin}, C.~L.},
        title = "{What is a cool-core cluster? a detailed analysis of the cores of the X-ray flux-limited HIFLUGCS cluster sample}",
      journal = {\aap},
         year = 2010,
        month = apr,
       volume = {513},
          eid = {A37},
        pages = {A37},
          doi = {10.1051/0004-6361/200912377},
archivePrefix = {arXiv},
       eprint = {0911.0409},
 primaryClass = {astro-ph.CO},
       adsurl = {https://ui.adsabs.harvard.edu/abs/2010A&A...513A..37H}
}

@ARTICLE{Ghirardini2019,
       author = {{Ghirardini}, V. and {Eckert}, D. and {Ettori}, S. and {Pointecouteau}, E. and {Molendi}, S. and {Gaspari}, M. and {Rossetti}, M. and {De Grandi}, S. and {Roncarelli}, M. and {Bourdin}, H. and {Mazzotta}, P. and {Rasia}, E. and {Vazza}, F.},
        title = "{Universal thermodynamic properties of the intracluster medium over two decades in radius in the X-COP sample}",
      journal = {\aap},
         year = 2019,
        month = jan,
       volume = {621},
          eid = {A41},
        pages = {A41},
          doi = {10.1051/0004-6361/201833325},
archivePrefix = {arXiv},
       eprint = {1805.00042},
 primaryClass = {astro-ph.CO},
       adsurl = {https://ui.adsabs.harvard.edu/abs/2019A&A...621A..41G}
}

@ARTICLE{Arnaud2010,
       author = {{Arnaud}, M. and {Pratt}, G.~W. and {Piffaretti}, R. and {B{\"o}hringer}, H. and {Croston}, J.~H. and {Pointecouteau}, E.},
        title = "{The universal galaxy cluster pressure profile from a representative sample of nearby systems (REXCESS) and the Y$_{SZ}$ - M$_{500}$ relation}",
      journal = {\aap},
         year = 2010,
        month = jul,
       volume = {517},
          eid = {A92},
        pages = {A92},
          doi = {10.1051/0004-6361/200913416},
archivePrefix = {arXiv},
       eprint = {0910.1234},
 primaryClass = {astro-ph.CO},
       adsurl = {https://ui.adsabs.harvard.edu/abs/2010A&A...517A..92A}
}

@ARTICLE{NFW1995,
       author = {{Navarro}, Julio F. and {Frenk}, Carlos S. and {White}, Simon D.~M.},
        title = "{Simulations of X-ray clusters}",
      journal = {\mnras},
         year = 1995,
        month = aug,
       volume = {275},
       number = {3},
        pages = {720-740},
          doi = {10.1093/mnras/275.3.720},
archivePrefix = {arXiv},
       eprint = {astro-ph/9408069},
 primaryClass = {astro-ph},
       adsurl = {https://ui.adsabs.harvard.edu/abs/1995MNRAS.275..720N}
}

@ARTICLE{NFW1996,
       author = {{Navarro}, Julio F. and {Frenk}, Carlos S. and {White}, Simon D.~M.},
        title = "{The Structure of Cold Dark Matter Halos}",
      journal = {\apj},
         year = 1996,
        month = may,
       volume = {462},
        pages = {563},
          doi = {10.1086/177173},
archivePrefix = {arXiv},
       eprint = {astro-ph/9508025},
 primaryClass = {astro-ph},
       adsurl = {https://ui.adsabs.harvard.edu/abs/1996ApJ...462..563N}
}

@ARTICLE{NFW1997,
       author = {{Navarro}, Julio F. and {Frenk}, Carlos S. and {White}, Simon D.~M.},
        title = "{A Universal Density Profile from Hierarchical Clustering}",
      journal = {\apj},
         year = 1997,
        month = dec,
       volume = {490},
       number = {2},
        pages = {493-508},
          doi = {10.1086/304888},
archivePrefix = {arXiv},
       eprint = {astro-ph/9611107},
 primaryClass = {astro-ph},
       adsurl = {https://ui.adsabs.harvard.edu/abs/1997ApJ...490..493N}
}

@ARTICLE{Hitomi2018,
       author = {{Hitomi Collaboration} and {Aharonian}, Felix and {Akamatsu}, Hiroki and {Akimoto}, Fumie and {Allen}, Steven W. and {Angelini}, Lorella and {Audard}, Marc and {Awaki}, Hisamitsu and {Axelsson}, Magnus and {Bamba}, Aya and {Bautz}, Marshall W. and {Blandford}, Roger and {Brenneman}, Laura W. and {Brown}, Gregory V. and {Bulbul}, Esra and {Cackett}, Edward M. and {Canning}, Rebecca E.~A. and {Chernyakova}, Maria and {Chiao}, Meng P. and {Coppi}, Paolo S. and {Costantini}, Elisa and {de Plaa}, Jelle and {de Vries}, Cor P. and {den Herder}, Jan-Willem and {Done}, Chris and {Dotani}, Tadayasu and {Ebisawa}, Ken and {Eckart}, Megan E. and {Enoto}, Teruaki and {Ezoe}, Yuichiro and {Fabian}, Andrew C. and {Ferrigno}, Carlo and {Foster}, Adam R. and {Fujimoto}, Ryuichi and {Fukazawa}, Yasushi and {Furuzawa}, Akihiro and {Galeazzi}, Massimiliano and {Gallo}, Luigi C. and {Gandhi}, Poshak and {Giustini}, Margherita and {Goldwurm}, Andrea and {Gu}, Liyi and {Guainazzi}, Matteo and {Haba}, Yoshito and {Hagino}, Kouichi and {Hamaguchi}, Kenji and {Harrus}, Ilana M. and {Hatsukade}, Isamu and {Hayashi}, Katsuhiro and {Hayashi}, Takayuki and {Hayashi}, Tasuku and {Hayashida}, Kiyoshi and {Hiraga}, Junko S. and {Hornschemeier}, Ann and {Hoshino}, Akio and {Hughes}, John P. and {Ichinohe}, Yuto and {Iizuka}, Ryo and {Inoue}, Hajime and {Inoue}, Shota and {Inoue}, Yoshiyuki and {Ishida}, Manabu and {Ishikawa}, Kumi and {Ishisaki}, Yoshitaka and {Iwai}, Masachika and {Kaastra}, Jelle and {Kallman}, Tim and {Kamae}, Tsuneyoshi and {Kataoka}, Jun and {Katsuda}, Satoru and {Kawai}, Nobuyuki and {Kelley}, Richard L. and {Kilbourne}, Caroline A. and {Kitaguchi}, Takao and {Kitamoto}, Shunji and {Kitayama}, Tetsu and {Kohmura}, Takayoshi and {Kokubun}, Motohide and {Koyama}, Katsuji and {Koyama}, Shu and {Kretschmar}, Peter and {Krimm}, Hans A. and {Kubota}, Aya and {Kunieda}, Hideyo and {Laurent}, Philippe and {Lee}, Shiu-Hang and {Leutenegger}, Maurice A. and {Limousin}, Olivier and {Loewenstein}, Michael and {Long}, Knox S. and {Lumb}, David and {Madejski}, Greg and {Maeda}, Yoshitomo and {Maier}, Daniel and {Makishima}, Kazuo and {Markevitch}, Maxim and {Matsumoto}, Hironori and {Matsushita}, Kyoko and {McCammon}, Dan and {McNamara}, Brian R. and {Mehdipour}, Missagh and {Miller}, Eric D. and {Miller}, Jon M. and {Mineshige}, Shin and {Mitsuda}, Kazuhisa and {Mitsuishi}, Ikuyuki and {Miyazawa}, Takuya and {Mizuno}, Tsunefumi and {Mori}, Hideyuki and {Mori}, Koji and {Mukai}, Koji and {Murakami}, Hiroshi and {Mushotzky}, Richard F. and {Nakagawa}, Takao and {Nakajima}, Hiroshi and {Nakamori}, Takeshi and {Nakashima}, Shinya and {Nakazawa}, Kazuhiro and {Nobukawa}, Kumiko K. and {Nobukawa}, Masayoshi and {Noda}, Hirofumi and {Odaka}, Hirokazu and {Ohashi}, Takaya and {Ohno}, Masanori and {Okajima}, Takashi and {Ota}, Naomi and {Ozaki}, Masanobu and {Paerels}, Frits and {Paltani}, St{\'e}phane and {Petre}, Robert and {Pinto}, Ciro and {Porter}, Frederick S. and {Pottschmidt}, Katja and {Reynolds}, Christopher S. and {Safi-Harb}, Samar and {Saito}, Shinya and {Sakai}, Kazuhiro and {Sasaki}, Toru and {Sato}, Goro and {Sato}, Kosuke and {Sato}, Rie and {Sawada}, Makoto and {Schartel}, Norbert and {Serlemtsos}, Peter J. and {Seta}, Hiromi and {Shidatsu}, Megumi and {Simionescu}, Aurora and {Smith}, Randall K. and {Soong}, Yang and {Stawarz}, {\L}ukasz and {Sugawara}, Yasuharu and {Sugita}, Satoshi and {Szymkowiak}, Andrew and {Tajima}, Hiroyasu and {Takahashi}, Hiromitsu and {Takahashi}, Tadayuki and {Takeda}, Shin'ichiro and {Takei}, Yoh and {Tamagawa}, Toru and {Tamura}, Takayuki and {Tanaka}, Keigo and {Tanaka}, Takaaki and {Tanaka}, Yasuo and {Tanaka}, Yasuyuki T. and {Tashiro}, Makoto S. and {Tawara}, Yuzuru and {Terada}, Yukikatsu and {Terashima}, Yuichi and {Tombesi}, Francesco and {Tomida}, Hiroshi and {Tsuboi}, Yohko and {Tsujimoto}, Masahiro and {Tsunemi}, Hiroshi and {Tsuru}, Takeshi Go and {Uchida}, Hiroyuki and {Uchiyama}, Hideki and {Uchiyama}, Yasunobu and {Ueda}, Shutaro and {Ueda}, Yoshihiro and {Uno}, Shin'ichiro and {Urry}, C. Megan and {Ursino}, Eugenio and {Wang}, Qian H.~S. and {Watanabe}, Shin and {Werner}, Norbert and {Wilkins}, Dan R. and {Williams}, Brian J. and {Yamada}, Shinya and {Yamaguchi}, Hiroya and {Yamaoka}, Kazutaka and {Yamasaki}, Noriko Y. and {Yamauchi}, Makoto and {Yamauchi}, Shigeo and {Yaqoob}, Tahir and {Yatsu}, Yoichi and {Yonetoku}, Daisuke and {Zhuravleva}, Irina and {Zoghbi}, Abderahmen},
        title = "{Atmospheric gas dynamics in the Perseus cluster observed with Hitomi}",
      journal = {\pasj},
         year = 2018,
        month = mar,
       volume = {70},
       number = {2},
          eid = {9},
        pages = {9},
          doi = {10.1093/pasj/psx138},
archivePrefix = {arXiv},
       eprint = {1711.00240},
 primaryClass = {astro-ph.HE},
       adsurl = {https://ui.adsabs.harvard.edu/abs/2018PASJ...70....9H}
}

@ARTICLE{TozziNorman2001,
       author = {{Tozzi}, Paolo and {Norman}, Colin},
        title = "{The Evolution of X-Ray Clusters and the Entropy of the Intracluster Medium}",
      journal = {\apj},
         year = 2001,
        month = jan,
       volume = {546},
       number = {1},
        pages = {63-84},
          doi = {10.1086/318237},
archivePrefix = {arXiv},
       eprint = {astro-ph/0003289},
 primaryClass = {astro-ph},
       adsurl = {https://ui.adsabs.harvard.edu/abs/2001ApJ...546...63T}
}

@ARTICLE{Pagagoulia2014,
       author = {{Panagoulia}, E.~K. and {Fabian}, A.~C. and {Sanders}, J.~S.},
        title = "{A volume-limited sample of X-ray galaxy groups and clusters - I. Radial entropy and cooling time profiles}",
      journal = {\mnras},
         year = 2014,
        month = mar,
       volume = {438},
       number = {3},
        pages = {2341-2354},
          doi = {10.1093/mnras/stt2349},
archivePrefix = {arXiv},
       eprint = {1312.0798},
 primaryClass = {astro-ph.CO},
       adsurl = {https://ui.adsabs.harvard.edu/abs/2014MNRAS.438.2341P}
}

@article{Sullivan2024b, title={Predicting the scaling relations between the dark matter halo mass and observables from generalised profiles II: Intracluster gas emission}, volume={41}, DOI={10.1017/pasa.2024.24}, journal={Publications of the Astronomical Society of Australia}, author={Sullivan, Andrew and Power, Chris and Bottrell, Connor and Robotham, Aaron and Shabala, Stanislav}, year={2024}, pages={e022}}

@ARTICLE{Pratt2019,
       author = {{Pratt}, G.~W. and {Arnaud}, M. and {Biviano}, A. and {Eckert}, D. and {Ettori}, S. and {Nagai}, D. and {Okabe}, N. and {Reiprich}, T.~H.},
        title = "{The Galaxy Cluster Mass Scale and Its Impact on Cosmological Constraints from the Cluster Population}",
      journal = {\ssr},
         year = 2019,
        month = feb,
       volume = {215},
       number = {2},
          eid = {25},
        pages = {25},
          doi = {10.1007/s11214-019-0591-0},
archivePrefix = {arXiv},
       eprint = {1902.10837},
 primaryClass = {astro-ph.CO},
       adsurl = {https://ui.adsabs.harvard.edu/abs/2019SSRv..215...25P}
}

@ARTICLE{Power2020,
       author = {{Power}, C. and {Elahi}, P.~J. and {Welker}, C. and {Knebe}, A. and {Pearce}, F.~R. and {Yepes}, G. and {Dav{\'e}}, R. and {Kay}, S.~T. and {McCarthy}, I.~G. and {Puchwein}, E. and {Borgani}, S. and {Cunnama}, D. and {Cui}, W. and {Schaye}, J.},
        title = "{NIFTY galaxy cluster simulations - VI. The dynamical imprint of substructure on gaseous cluster outskirts}",
      journal = {\mnras},
         year = 2020,
        month = jan,
       volume = {491},
       number = {3},
        pages = {3923-3936},
          doi = {10.1093/mnras/stz3176},
archivePrefix = {arXiv},
       eprint = {1810.00534},
 primaryClass = {astro-ph.CO},
       adsurl = {https://ui.adsabs.harvard.edu/abs/2020MNRAS.491.3923P}
}

@ARTICLE{Shabala2018,
       author = {{Shabala}, Stanislav S.},
        title = "{The role of environment in the observed Fundamental Plane of radio active galactic nuclei}",
      journal = {\mnras},
         year = 2018,
        month = aug,
       volume = {478},
       number = {4},
        pages = {5074-5080},
          doi = {10.1093/mnras/sty1328},
archivePrefix = {arXiv},
       eprint = {1805.06600},
 primaryClass = {astro-ph.GA},
       adsurl = {https://ui.adsabs.harvard.edu/abs/2018MNRAS.478.5074S}
}

@ARTICLE{Yates2018,
       author = {{Yates}, Patrick M. and {Shabala}, Stanislav S. and {Krause}, Martin G.~H.},
        title = "{Observability of intermittent radio sources in galaxy groups and clusters}",
      journal = {\mnras},
         year = 2018,
        month = nov,
       volume = {480},
       number = {4},
        pages = {5286-5306},
          doi = {10.1093/mnras/sty2191},
archivePrefix = {arXiv},
       eprint = {1808.03026},
 primaryClass = {astro-ph.HE},
       adsurl = {https://ui.adsabs.harvard.edu/abs/2018MNRAS.480.5286Y}
}

@ARTICLE{Salvati2019,
       author = {{Salvati}, Laura and {Douspis}, Marian and {Ritz}, Anna and {Aghanim}, Nabila and {Babul}, Arif},
        title = "{Mass bias evolution in tSZ cluster cosmology}",
      journal = {\aap},
         year = 2019,
        month = jun,
       volume = {626},
          eid = {A27},
        pages = {A27},
          doi = {10.1051/0004-6361/201935041},
archivePrefix = {arXiv},
       eprint = {1901.03096},
 primaryClass = {astro-ph.CO},
       adsurl = {https://ui.adsabs.harvard.edu/abs/2019A&A...626A..27S}
}

@ARTICLE{Hitomi2016,
       author = {{Hitomi Collaboration} and {Aharonian}, Felix and {Akamatsu}, Hiroki and {Akimoto}, Fumie and {Allen}, Steven W. and {Anabuki}, Naohisa and {Angelini}, Lorella and {Arnaud}, Keith and {Audard}, Marc and {Awaki}, Hisamitsu and {Axelsson}, Magnus and {Bamba}, Aya and {Bautz}, Marshall and {Blandford}, Roger and {Brenneman}, Laura and {Brown}, Gregory V. and {Bulbul}, Esra and {Cackett}, Edward and {Chernyakova}, Maria and {Chiao}, Meng and {Coppi}, Paolo and {Costantini}, Elisa and {de Plaa}, Jelle and {den Herder}, Jan-Willem and {Done}, Chris and {Dotani}, Tadayasu and {Ebisawa}, Ken and {Eckart}, Megan and {Enoto}, Teruaki and {Ezoe}, Yuichiro and {Fabian}, Andrew C. and {Ferrigno}, Carlo and {Foster}, Adam and {Fujimoto}, Ryuichi and {Fukazawa}, Yasushi and {Furuzawa}, Akihiro and {Galeazzi}, Massimiliano and {Gallo}, Luigi and {Gandhi}, Poshak and {Giustini}, Margherita and {Goldwurm}, Andrea and {Gu}, Liyi and {Guainazzi}, Matteo and {Haba}, Yoshito and {Hagino}, Kouichi and {Hamaguchi}, Kenji and {Harrus}, Ilana and {Hatsukade}, Isamu and {Hayashi}, Katsuhiro and {Hayashi}, Takayuki and {Hayashida}, Kiyoshi and {Hiraga}, Junko and {Hornschemeier}, Ann and {Hoshino}, Akio and {Hughes}, John and {Iizuka}, Ryo and {Inoue}, Hajime and {Inoue}, Yoshiyuki and {Ishibashi}, Kazunori and {Ishida}, Manabu and {Ishikawa}, Kumi and {Ishisaki}, Yoshitaka and {Itoh}, Masayuki and {Iyomoto}, Naoko and {Kaastra}, Jelle and {Kallman}, Timothy and {Kamae}, Tuneyoshi and {Kara}, Erin and {Kataoka}, Jun and {Katsuda}, Satoru and {Katsuta}, Junichiro and {Kawaharada}, Madoka and {Kawai}, Nobuyuki and {Kelley}, Richard and {Khangulyan}, Dmitry and {Kilbourne}, Caroline and {King}, Ashley and {Kitaguchi}, Takao and {Kitamoto}, Shunji and {Kitayama}, Tetsu and {Kohmura}, Takayoshi and {Kokubun}, Motohide and {Koyama}, Shu and {Koyama}, Katsuji and {Kretschmar}, Peter and {Krimm}, Hans and {Kubota}, Aya and {Kunieda}, Hideyo and {Laurent}, Philippe and {Lebrun}, Fran{\c{c}}ois and {Lee}, Shiu-Hang and {Leutenegger}, Maurice and {Limousin}, Olivier and {Loewenstein}, Michael and {Long}, Knox S. and {Lumb}, David and {Madejski}, Grzegorz and {Maeda}, Yoshitomo and {Maier}, Daniel and {Makishima}, Kazuo and {Markevitch}, Maxim and {Matsumoto}, Hironori and {Matsushita}, Kyoko and {McCammon}, Dan and {McNamara}, Brian and {Mehdipour}, Missagh and {Miller}, Eric and {Miller}, Jon and {Mineshige}, Shin and {Mitsuda}, Kazuhisa and {Mitsuishi}, Ikuyuki and {Miyazawa}, Takuya and {Mizuno}, Tsunefumi and {Mori}, Hideyuki and {Mori}, Koji and {Moseley}, Harvey and {Mukai}, Koji and {Murakami}, Hiroshi and {Murakami}, Toshio and {Mushotzky}, Richard and {Nagino}, Ryo and {Nakagawa}, Takao and {Nakajima}, Hiroshi and {Nakamori}, Takeshi and {Nakano}, Toshio and {Nakashima}, Shinya and {Nakazawa}, Kazuhiro and {Nobukawa}, Masayoshi and {Noda}, Hirofumi and {Nomachi}, Masaharu and {O'Dell}, Steve and {Odaka}, Hirokazu and {Ohashi}, Takaya and {Ohno}, Masanori and {Okajima}, Takashi and {Ota}, Naomi and {Ozaki}, Masanobu and {Paerels}, Frits and {Paltani}, Stephane and {Parmar}, Arvind and {Petre}, Robert and {Pinto}, Ciro and {Pohl}, Martin and {Porter}, F. Scott and {Pottschmidt}, Katja and {Ramsey}, Brian and {Reynolds}, Christopher and {Russell}, Helen and {Safi-Harb}, Samar and {Saito}, Shinya and {Sakai}, Kazuhiro and {Sameshima}, Hiroaki and {Sato}, Goro and {Sato}, Kosuke and {Sato}, Rie and {Sawada}, Makoto and {Schartel}, Norbert and {Serlemitsos}, Peter and {Seta}, Hiromi and {Shidatsu}, Megumi and {Simionescu}, Aurora and {Smith}, Randall and {Soong}, Yang and {Stawarz}, Lukasz and {Sugawara}, Yasuharu and {Sugita}, Satoshi and {Szymkowiak}, Andrew and {Tajima}, Hiroyasu and {Takahashi}, Hiromitsu and {Takahashi}, Tadayuki and {Takeda}, Shin'Ichiro and {Takei}, Yoh and {Tamagawa}, Toru and {Tamura}, Keisuke and {Tamura}, Takayuki and {Tanaka}, Takaaki and {Tanaka}, Yasuo and {Tanaka}, Yasuyuki and {Tashiro}, Makoto and {Tawara}, Yuzuru and {Terada}, Yukikatsu and {Terashima}, Yuichi and {Tombesi}, Francesco and {Tomida}, Hiroshi and {Tsuboi}, Yohko and {Tsujimoto}, Masahiro and {Tsunemi}, Hiroshi and {Tsuru}, Takeshi and {Uchida}, Hiroyuki and {Uchiyama}, Hideki and {Uchiyama}, Yasunobu and {Ueda}, Shutaro and {Ueda}, Yoshihiro and {Ueno}, Shiro and {Uno}, Shin'Ichiro and {Urry}, Meg and {Ursino}, Eugenio and {de Vries}, Cor and {Watanabe}, Shin and {Werner}, Norbert and {Wik}, Daniel and {Wilkins}, Dan and {Williams}, Brian and {Yamada}, Shinya and {Yamaguchi}, Hiroya and {Yamaoka}, Kazutaka and {Yamasaki}, Noriko Y. and {Yamauchi}, Makoto and {Yamauchi}, Shigeo and {Yaqoob}, Tahir and {Yatsu}, Yoichi and {Yonetoku}, Daisuke and {Yoshida}, Atsumasa and {Yuasa}, Takayuki and {Zhuravleva}, Irina and {Zoghbi}, Abderahmen},
        title = "{The quiescent intracluster medium in the core of the Perseus cluster}",
      journal = {\nat},
         year = 2016,
        month = jul,
       volume = {535},
       number = {7610},
        pages = {117-121},
          doi = {10.1038/nature18627},
archivePrefix = {arXiv},
       eprint = {1607.04487},
 primaryClass = {astro-ph.GA},
       adsurl = {https://ui.adsabs.harvard.edu/abs/2016Natur.535..117H}
}

@ARTICLE{Dupourque2023,
       author = {{Dupourqu{\'e}}, S. and {Clerc}, N. and {Pointecouteau}, E. and {Eckert}, D. and {Ettori}, S. and {Vazza}, F.},
        title = "{Investigating the turbulent hot gas in X-COP galaxy clusters}",
      journal = {\aap},
         year = 2023,
        month = may,
       volume = {673},
          eid = {A91},
        pages = {A91},
          doi = {10.1051/0004-6361/202245779},
archivePrefix = {arXiv},
       eprint = {2303.15102},
 primaryClass = {astro-ph.CO},
       adsurl = {https://ui.adsabs.harvard.edu/abs/2023A&A...673A..91D}
}

@ARTICLE{Campitiello2022,
       author = {{Campitiello}, M.~G. and {Ettori}, S. and {Lovisari}, L. and {Bartalucci}, I. and {Eckert}, D. and {Rasia}, E. and {Rossetti}, M. and {Gastaldello}, F. and {Pratt}, G.~W. and {Maughan}, B. and {Pointecouteau}, E. and {Sereno}, M. and {Biffi}, V. and {Borgani}, S. and {De Luca}, F. and {De Petris}, M. and {Gaspari}, M. and {Ghizzardi}, S. and {Mazzotta}, P. and {Molendi}, S.},
        title = "{CHEX-MATE: Morphological analysis of the sample}",
      journal = {\aap},
         year = 2022,
        month = sep,
       volume = {665},
          eid = {A117},
        pages = {A117},
          doi = {10.1051/0004-6361/202243470},
archivePrefix = {arXiv},
       eprint = {2205.11326},
 primaryClass = {astro-ph.CO},
       adsurl = {https://ui.adsabs.harvard.edu/abs/2022A&A...665A.117C}
}

\end{document}